\documentclass[preprint2]{aastex}

\newcommand{\be}{\begin{equation}}
\newcommand{\ee}{\end{equation}}
\newcommand{\beq}{\begin{eqnarray}}
\newcommand{\eeq}{\end{eqnarray}}

\slugcomment{Version of \today}

\shorttitle{Loop Temperature Profiles for Non-uniform Heating}
\shortauthors{Martens}

\begin{document}

\title{Scaling Laws and Temperature Profiles for Solar and Stellar Coronal Loops with Non-uniform
Heating}
\author{P.C.H. Martens\altaffilmark{1}}
\affil{Harvard-Smithsonian Center for Astrophysics,  Cambridge, MA 02138}
\altaffiltext{1}{Affiliate Professor, Physics Department, Montana State University -- Bozeman}
\email{pmartens@cfa.harvard.edu}

\begin{abstract}

The bulk of solar coronal radiative loss consists of soft X-ray emission from quasi-static
loops at the cores of Active Regions. In order to develop diagnostics for determining the
heating mechanism of these loops from observations by coronal 
imaging instruments, I have developed analytical solutions
for the temperature structure and scaling laws of loop strands for a
wide range of heating functions, including footpoint heating,
uniform heating, and heating concentrated at the loop apex.
Key results are that the temperature profile depends only weakly on the heating
distribution -- not sufficiently to be of significant diagnostic value --  and that the scaling laws
survive for this wide range of heating distributions, but with the constant of proportionality
in the RTV scaling law ($P_{0}L\, \thicksim\, T_{max}^3$) depending on the specific heating
function.  Furthermore, quasi-static analytical solutions do not exist for an excessive concentration
of heating near the loop footpoints, a result in agreement with recent numerical simulations.
It is demonstrated that a generalization of the solutions to the case of a strand with a variable
diameter leads to only relatively small correction factors in the scaling laws and
temperature profiles for constant diameter loop strands.  A quintet of leading theoretical
coronal heating mechanisms is shown to be captured by the formalism of this paper, and
the differences in thermal structure between them may be verified through observations.
Preliminary results from full numerical simulations demonstrate that, despite the simplifying
assumptions, the analytical solutions from this paper are stable and accurate.

 \end{abstract}



\section{Introduction}

The emission from the solar corona in the Extreme Ultra-Violet (EUV) and in X-rays
has been observed since the advent of space-borne telescopes.   The mere existence
of that emission demonstrates that the solar corona has a very high temperature and
therefore is not in radiative equilibrium with the underlying photosphere.   The solar
corona requires energy deposition of some sort to maintain its high temperature, and it
is now widely assumed that this coronal heating is magnetic in nature, either through
the dissipation of electrical currents or of incoming MHD waves.

The S-054 X-ray telescope (Vaiana et al.\  1973) on the {\it Skylab} mission in particular
has contributed greatly to elucidating the nature of the coronal high temperature emission. 
It became clear that the
emission from the corona is highly structured in so-called coronal loops; elongated
bent high aspect ratio cylinders that appear to be rooted in the solar chromosphere at
both ends.  The  {\it Skylab} observations further showed that the footpoints of the X-ray
emitting coronal
loops are predominantly located at photospheric plage regions or the penumbrae --
but not at the umbrae -- of Sunspots.  This demonstrates a clear connection with
enhanced photospheric magnetic field, but curiously not with the strongest  field.
After {\it Skylab}  soft X-ray images from the solar corona have been extensively studied
by the Soft X-ray Telescope (SXT) onboard {\it Yohkoh} (Tsuneta et al.\ 1991)
\nocite{tsuneta-etal91} and currently by the X-ray Telescope (XRT, Golub et al.\ 2007)
\nocite{golub-etal07}
onboard the follow-up Japanese solar mission, {\it Hinode}.  The
latter has the unprecedented resolution in X-rays of two arcseconds, a factor of 2.5 or
more better than its predecessors.

While the coronal soft  X-ray emission originates from loop plasma at typical
temperatures of several (2-5) MK, solar EUV emission comes from plasma with a
temperature of 1-2 MK.    The EUV full disk corona has been observed for an entire
solar cycle now by the Extreme Ultraviolet Telescope (EIT, Delaboudiniere et al.\ 
1995) onboard the  {\it Solar and Heliospheric Observatory} ({\it SoHO}),
\nocite{delaboudiniere-etal95} and since 1998 with the
unparalleled resolution of one arcsecond by the {\it Transition Region and Coronal
Explorer} ({\it TRACE}, Handy et al.\ 1999).  \nocite{handy-etal99}  The {\it TRACE}
observations in particular show that the EUV corona at $\thicksim$ 1 MK (in the
171 \AA\  passband) is clearly structured, with much more pronounced loops than 
the hotter corona, as observed in narrow-band images, for example, in the {\it TRACE}
285 \AA\ passband.  The same appears, interestingly enough, not to be so in
broadband coronal soft X-ray images, such as those from XRT.
I point out here that the {\it TRACE} images are usually
shown on a linear intensity scale after subtraction of the dark current, while soft
X-ray images, such as those from SXT and XRT, usually show the square root of
the image intensity or the natural logarithm, also after subtraction of a pedestal.
The latter rendition is intended to bring out the fainter structures more clearly,
motivated by the fact that in soft X-rays coronal emission tends to be dominated by
a few very bright loops, i.e., the dynamic range in intensity of the
emitting structures is much larger and differently distributed than in the EUV. 
This manner of image rendering also has the unintended effect
of enhancing the background in soft X-ray images in comparison with EUV images,
adding to the appearance of more ``fuzzy" coronal images that is not real.  In fact,
recent work by Cirtain et al.\ (2006) \nocite{cirtain-etal06} demonstrates that in typical
Active Regions more than three quarters of the emission in the EUV originates from
the unresolved {\it TRACE} background (i.e., non-loop emission), while inspection of
the {\it Hinode} soft X-ray images yields a fraction of a fifth at the very most
(Weber 2008, private communication).

The relevance of the above is that the soft X-ray and EUV emission from the solar corona
is completely dominated by that of solar Active Regions;  Aschwanden (2001) and
Aschwanden et al.\  (2007) estimate that the Active Regions account for 82.4 \% of the
heating requirement of the corona. \nocite{aschwanden-etal07, aschwanden01}
Therefore, in order to understand the origin of solar coronal heating one needs to consider
the heating of Active Regions (ARs), and in particular the heating of soft X-ray emitting 
AR loops, which dominate the energy output.

\begin{figure*}[!t]
\epsscale{1.80}
\plotone{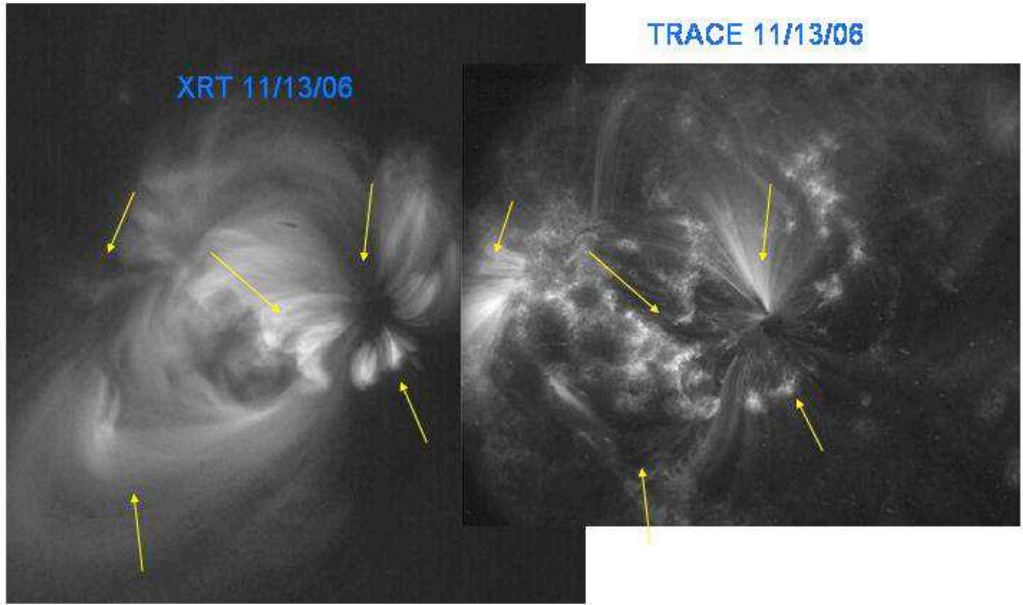}
\caption{Simultaneous (within one sec) images of an Active Region (AR) with the XRT Thin Be
filter, and the {\it TRACE} 171 \AA\ filter.  The arrows, at identical locations in both images, point to,
from left to right and top to bottom:  a) a fan of EUV loops in the periphery of the AR, b) hot Soft X-ray
loops in the core of the AR, with EUV moss at their footpoints, c) a second fan of EUV loops from near
the umbra of a sunspot, with no corresponding soft X-ray emission, d) diffuse hot and very long
Soft X-ray loops surrounding the core, with no corresponding EUV emission, and e) short bright soft
X-ray loops with corresponding EUV moss.  Image courtesy of Dr.\ Leon Golub.
\label{fig:euv-sxr}}
\end{figure*}

Typically one will find soft X-ray loops in the center of ARs, and the somewhat cooler
and often longer
EUV loops in the periphery.  In addition there is substantial EUV emission from the
so-called ``moss",  located at and just above the footpoints of the hotter soft X-ray loops.
This is illustrated by Figure~\ref{fig:euv-sxr}, showing two simultaneous and co-aligned
coronal images, one in the EUV taken by {\it TRACE}  and one in soft X-rays taken by
XRT on {\it Hinode}.

The hotter soft X-ray loops typically appear relatively steady over time-scales that are
longer than their radiative and conductive cooling times, while the EUV loops seem more
dynamic, often
exhibiting a cooling collapse as well as jets and siphon flows of chromospheric material
entering these loops.  The absence of significant EUV emission, except for the moss, in
the soft X-ray loops at the centers of ARs led  Antiochos et al.\ (2003) 
\nocite{antiochos-etal03}
to the conclusion that these loops must be either steadily heated, or intermittently --
like, say, in nano-flares -- with a repeat time that is shorter than the loop cooling time:
if that weren't the case these loops should be observed emitting in EUV bands as they
cool down.  Recent work by Winebarger, Warren, \& Falconer (2008) confirms this
result.  \nocite{winebarger-warren-falconer08}
Antiochos et al.\  quantify their result with numerical simulations of loop
hydrodynamics using the sophisticated NRL loop code.   They further
demonstrate that since the observed moss emission from the footpoints of the
soft X-ray emitting loops varies little, of the order of 10\% over hours, the coronal
loop heating mechanism must be rather steady in time.  This justifies the quasi-static
modeling of the thermal structure of the AR soft X-ray loops, as will be done in the
present paper.  Quasi-static is defined here as having the constraints on the heating
rate mentioned above, and having plasma-flows in the loops that are substantially
subsonic (say, less than 1/3 of the speed of sound).  The analysis of recent observations
with XRT and EIT on {\it Hinode} by Warren et al.\ (2007) confirms this picture by showing
a steady output for the bulk of the emission from the AR analyzed, supplemented by
transient brightenings in a limited number of loops.
\nocite{warren-etal07} 

It was mentioned above that the EUV loops in the periphery of ARs appear much
more dynamic.  Yet Aschwanden, Schrijver, \& Alexander (2001) find from the  
\nocite{aschwanden-schrijver-alexander01}
analysis of some 40 loops observed by {\it TRACE} that about 30\% of these EUV
loops appear in quasi-steady hydrostatic equilibrium.  For the remainder that is
dynamic one can show that a
quasi-static model is a good zeroth order approximation for the thermal structure
as long as any plasma motions are substantially subsonic, as defined above.    The
reason for that is that the kinetic terms in the energy equations scale as the square
of the ratio of the plasma velocity to the sound speed.  Therefore, also for the more
dynamic EUV AR loops quasi-static modeling is a useful first approach, certainly if
one can obtain general analytical results as I will do in this paper.

Having argued that coronal heating is concentrated in soft X-ray emitting AR loops,
and that for these loops quasi-static modeling of the thermal structure is an acceptable
approach, I will derive in the present paper full analytical solutions for the thermal structure
of coronal loops based upon these premises.   Analytical solutions and approximations 
have been derived by many since the pioneering papers by Rosner, Tucker, \& Vaiana
(1978b) and Craig, McClymont, \& Underwood (1978)
\nocite{rtv78,craig-mcclymont-underwood78}
but full analytical solutions have been limited to the case of uniform quasi-static heating of
loops.  Numerical simulations, on the other hand,
have investigated non-uniform heating, in particular cases where heating is concentrated
near the loop footpoints (e.g.,  Serio et al.\ 1981; Aschwanden et al.\ 2001;  M\"{u}ller,
Hansteen, \& Peter 2003; M\"{u}ller, Peter, \& Hansteen 2004; M\"{u}ller et al.\ 2005). 
\nocite{aschwanden-schrijver-alexander01, serio-etal81, mueller-hansteen-peter03,
mueller-peter-hansteen04, mueller-etal05}
Here I will derive
analytical solutions for the loop temperature profile and scaling laws for non-uniform
heating functions, including cases in which the heating is concentrated near the
footpoints and around the looptop.  These solutions produce generalizations of
the now 30 year old RTV scaling laws for uniform heating that have some surprising
properties that, as I will show, may be used as a diagnostic for determining the heating
profile in the loops.   On the other hand, it will be demonstrated that the temperature
profile of coronal loops is rather insensitive to the distribution of the coronal heating, so
that even very precise measurements of the temperature profile along loops would
not be very useful for determining the heating distribution.

In modeling coronal loops it is usually assumed that the loop has a uniform cross-section
along its length.  This is what observations, both in EUV and soft X-rays, indicate to be the
case (Klimchuk 2000, Watko \& Klimchuk 2000). \nocite{klimchuk00, watko-klimchuk00}
Visual inspection of Figure~\ref{fig:euv-sxr}
seems to confirm this.  However, a recent paper by DeForest (2007) \nocite{deforest07} has
cast doubt upon this notion.  To understand the argument I first need to introduce the concept
of ``strands",  a term that will be used throughout this paper.  Conduction across the magnetic
field is very ineffective in coronal plasma.  The basic reason is that the mean free path for
perpendicular particle motions is limited to the ion gyro radius, which is only of the order of
centimeters, while the mean free path along the magnetic field can be hundreds of kilometers
or more for thermal electrons.  Hence conductive energy transport along the magnetic field
is far more effective than across the field.   That results in the situation, already
recognized in Rosner et al.\ (1978b) and used ever since in loop modeling,
that the solar corona is divided in independent mini-atmospheres, or ``strands",  stretching
out along magnetic field lines from one chromospheric footpoint to the other, but with
very tiny cross-sections, possibly down to an ion gyro-radius.  These strands are then
modeled as one-dimensional structures, with the single coordinate along a field line. 

The loops observed by solar imagers, from the {\it Skylab} telescopes to {\it TRACE} and
XRT on {\it Hinode}, although having aspect ratios of 10\% and smaller, still have radii
of several thousands of km, down to perhaps 1000 km for the thinnest {\it TRACE} loops. 
Therefore these observed loops may consist of many, up to thousands, of individual
``strands".  Note that it is interesting, and perhaps a clue to the physical mechanism of
the heating mechanism, that these ``stands" tend to cluster together in the
much wider loops that are observed -- there is no a priori reason strands could not be 
distributed evenly over ARs, thereby creating a much smoother appearance.  The 
many strands within an observed loop can all  have different heating rates and
temperature structures.  On the
other hand, it is possible that the observed loops consist of many strands that are heated
in an identical manner and therefore all have the same temperature structure.   It was
pointed out by Martens, Cirtain, \& Schmelz (2002) \nocite{martens-cirtain-schmelz02}
that if the former is true one would expect that pixels from these loops have plasma from
different strands at multiple temperatures contributing to the emission, and therefore that
the Differential Emission Measures (DEMs, Pottasch 1963) \nocite{pottasch63}
for these pixels would be broad. 
Martens et al.\ then showed that for broad loop DEMs the customary method
of determining temperatures from filter ratios (for narrow-band images) and line-ratios (for
spectrometers) leads to misleading results.  They demonstrated that for broad DEMs the
ratio for one pair of lines or passbands in general produces a different temperature than
those from another pair, exactly what has been observed (e.g., Schmelz et al.\ 2001;
Reale et al.\ 2007). \nocite{schmelz-etal01, reale-etal07}  Weber et al.\ (2005) 
carried this approach further through simulations and found just how narrow a DEM has
to be for the filter or line-ratio method to yield a single, correct, filter-ratio temperature.
\nocite{weber-etal05}  In general these results imply one cannot assign a single
temperature to each pixel within a multi-strand multi-temperature loop.

The answer to the question whether observed loops are multi-thermal or isothermal,
obviously depends on the resolution of the telescope that is used: once we get down
in resolution to the size of the elementary strands, the answer must be isothermal.
However, so-far the answer for Soft X-ray loops observed by the currently highest
resolution soft X-ray telescope, XRT, is multi-thermal (Reale et al.\ 2007), and therefore
the observations can only be matched by models that contain a multitude of elementary
strands.  This leaves the modeler with a large number of parameters to adjust (number of
strands and heating rate and profile in each) which implies analytical solutions for each
individual strand -- as will be derived in this paper -- are very useful, since it makes it much
less computationally demanding to explore the parameter space of the models in trying
to obtain a forward fit to the data.  For EUV loops there is a heated debate on whether
the thinnest loops observed by {\it TRACE} are isothermal (e.g., Aschwanden \&
Nightingale 2005) \nocite{aschwanden-nightingale05}
or multi-thermal (e.g., Martens et al.\  2002; Schmelz \& Martens 2006; Schmelz,
Kashyap, \& Weber 2007), an issue that I will not further address in this paper.
\nocite{schmelz-martens06, schmelz-kashyap-weber07}
The observations by the Atmospheric Imaging Assembly (AIA)  onboard SDO, to be
launched by the end of 2008, might resolve this question for the {\it TRACE} loops, since
AIA has the same spatial
resolution as {\it TRACE}, but a much better spectral resolution due to its larger number of
EUV passbands.  Regardless of the outcome of the debate, it would be a great step
forward when the first isothermal strands are detected, because that puts strong
constraints on the physics of the heating mechanism.

As an aside, from the considerations above it also follows that the loops observed in
the EUV and soft X-rays also outline the magnetic field, and therefore are a great diagnostic
for the structure and topology of the coronal magnetic field, as the high resolution {\it TRACE}
images so beautifully demonstrate.

Now let's return to the issue of the cross-section of loops and strands.  In order to model a
strand one has to make an assumption about the variation of its cross-section, i.e., is it
constant or does the cross-section increase from footpoint to top?  As was stated above,
observed loops tend to show a constant cross-section, but that is not necessarily true for
the strands within them, as pointed out by DeForest (2007).   Extrapolations of the coronal
magnetic field from the photospheric boundary nearly always yield flux tubes expanding with
height, so why is it that observed loops do not follow that trend?  DeForest shows a clear
example of a polar plume observed by EIT (his Fig.~3) that appears to have a constant
diameter, but the same plume observed by {\it TRACE} at a five times higher resolution,
consists of a number of smaller strands, while the whole of the plume clearly shows an
increasing cross-section from the bottom up.
The analysis by DeForest demonstrates that near the limit of its
resolution any telescope will show strands that are expanding as having a constant
diameter, because of the smearing inherent in the Point Spread Function of the telescope,
because of photon noise, and because of the background and foreground emission along
the line-of-sight.  DeForest then shows that if these strands are in fact expanding this can
explain why the observed emission measure at the top of loops in {\it TRACE} is 
typically larger than would
be expected from modeling constant cross-section strands -- simply because there is more
emitting volume near the top. I should point out here
that others (e.g., Lopez-Fuentes, Demoulin, \& Klimchuk 2008) strongly disagree with
DeForest's results, arguing, based upon their simulations, that even for loops 
that are close to the telescope's resolution the diameter can be correctly measured.
\nocite{lopez-fuentes-demoulin-klimchuk08, deforest07}  However, the authors offer no
suggestion how to reconcile the observations of constant loop diameter with magnetic
filed extrapolations almost invariably showing the opposite.  (To be fair, that is outside
the scope of their paper).

This issue not being resolved yet, I have developed analytical solutions for strands both
with constant and varying diameters.  It is shown that the initial solution derived for
constant cross-section strands can be fairly easily generalized to a class of non-constant
diameter solutions, and, most importantly, that the resulting temperature profiles and
scaling laws only differ by a small correction factor from the uniform diameter case.  This
fortunate coincidence makes it possible to produce forward folding models for multi-strand
loops without having to choose between uniform or varying strand diameter, while at the
same time this result allows for an observational verification of the degree of uniformity
of the strand diameter, because the emitting volume within any loop varies proportional
to the change in the strand cross-section, as already analyzed  by DeForest (2007).

In summary of this introduction:  in this paper I will develop analytical solutions for the
temperature profile and scaling laws of the strands that make up the AR solar coronal
loops observed by EUV and soft X-ray telescopes.  The emission of these loops represents
the bulk of the solar coronal emission, and the comparison of their thermal structure with
the results from analytical and numerical analyses can therefore contribute to unveiling the 
detailed physics of the coronal heating mechanism.  

In \S~\ref{sec:formalism} I will first introduce and validate the approximations made in
rendering the equations for the thermal structure of strands amenable to analytical treatment.  I will
then derive explicitly solutions that apply to a whole range of non-uniform heating functions, from
footpoint heating to apex heating.  I will present both the standard case of a uniformly increasing
temperature from base to apex, as well as solutions with temperature reversals.

In \S~\ref{sec:interpretation} I will present a physical analysis of the results, and in particular
consider their applicability as a diagnostic for the heating mechanism of strands.  To illustrate this
I will formulate five often considered coronal heating mechanisms in terms of the expression
used for the heating function in this paper, and compare the observable predictions for the
resulting  temperature profiles and scaling laws.  In \S~\ref{sec:variable} I
will generalize the solutions for the temperature structure and scaling laws from the case of a
uniform loop cross-section to the case of a varying one, in particular the situation of a gradually
expanding loop diameter in going from footpoint to looptop.  Finally, in \S~\ref{sec:multiple-t} I
will analyze the solutions exhibiting temperature reversals, that, I will argue, apply to prominences.

In the Discussion section I will conclude by describing the heritage of the mathematical
description used in this paper, and comment on the severity of the approximations used in the
analytical treatment as well as on the existence, stability, and accuracy of the solutions.

\section{Formalism \label{sec:formalism}}

The classical derivation of the RTV scaling laws (Rosner et al.\ 1978b) \nocite{rtv78}
uses a coronal heating function that is a constant in space and time as well as constant pressure throughout the loop.  The latter assumption is reasonable when the height of the top of the loop above the chromosphere is less than the pressure scale height for the peak temperature.  Landini \& Monsignori-Fossi (1975), 
\nocite{landini-monsignori-fossi75}
followed by many others including Martens et al.\  (2000), \nocite{martens-kankelborg-berger00}
approximate the radiative loss function for constant pressure by a single power-law.  The latter then explicitly demonstrated that this approximation is quite accurate for the relevant temperature range of the corona and
transition region.  
Kuin \& Martens (1982) showed that this leads to particularly simple analytical expressions for the temperature 
profile in an isobaric and uniformly heated coronal loop and to exact expressions for the two scaling laws.

I will now derive equally simple results for a parametrised non-uniform heating function that, depending upon the choice of the parameters, can represent heating concentrated at the footpoints of the loop, at the apex, and heating that is constant. I will show that while the form of the first RTV scaling law (P$_0$L $\thicksim$ T$_{max}^3$, with P$_0$ the loop pressure, L, the half-length of the loop, and T$_{max}$ the maximum temperature at the apex) is preserved for all these heating functions, the value of the constant of proportionality depends on the spatial distribution of the heating.  This constant becomes larger as heating becomes more concentrated at the footpoints, so that for the same T$_{max}$ and L the loop pressure and hence average density are larger than for the case of constant heating.  Therefore these loops are, in fact, not overdense with respect to the scaling laws, only overdense with respect to the scaling law for uniform heating.

\subsection{Uniformly Increasing Temperature Solution  \label{sec:t-solution}}

For a static coronal loop under these assumptions the momentum equation yields constant pressure and the energy equation reduces to (e.g., Martens et al.\ 2000) \nocite{martens-kankelborg-berger00}
\be
\label{eq:balance} \frac{d}{dz}(\kappa_{0}T^{5/2}\frac{dT}{dz})\,+
\,E_{h}\,-\,P_{0}^2\chi_{0}T^{-(2+\gamma)}\,=\,0.
\ee
Here z is the coordinate along the axis of  the strand, measured from the base of the transition region, 
T(z) is the temperature profile, E$_h$ the unknown heating function, and $\chi_0$ and $\gamma$ the parametrisation constants in the power-law approximation of the radiative loss function for constant pressure.  Using the Feldman (1992) \nocite{feldman92} abundances that apply to closed coronal regions, Martens et al.\ (2000) derive that good values are $\gamma$ = 0.5, and $\chi_0$ = 10$^{12.41}$ in cgs units. The remaining term in this equation represents conductive heating or cooling, with  $\kappa_{0}$ a constant of value $1.1\times10^{-6}\,$erg$\,$cm$^{-1}\,$K$^{-1}$ (Chapman \& Cowling 1939; Spitzer 1962).
 \nocite{chapman-cowling39, spitzer62}.

Rather than assuming uniform heating I now use a heating function that has a power-law dependence upon density and temperature E$_h\, \thicksim\,  \rho^{\beta}\, T^{(\alpha+\beta)}$, which, as I will demonstrate below, can plausibly represent a number of physical coronal heating mechanisms.  Using the gas-law to eliminate the density in favor of the constant pressure one finds
\be
E_h\ =\ H\, P_0^{\beta}T^{\alpha},
\label{eq:heatingf}
\ee
where H is a constant of proportionality.  Clearly the double power-law dependence upon density and heating collapses to a single power-law for constant pressure after using the gas law, but this is not the case when the pressure scale height becomes comparable to the loop height.  Uniform heating is represented by the parameter choice $\alpha$=$\beta$=0, while for a uniformly increasing temperature from footpoint to loop apex, negative $\alpha$'s represent footpoint heating, and positive $\alpha$'s apex heating.  This formalism for non-uniform coronal heating was introduced by Craig, McClymont, \& Underwood (1978), and applied by
Landini \& Monsignori-Fossi (1981), Torricelli-Ciamponi, Einaudi, \& Chiuderi, and Kuin \& Martens (1982).
Bray et al.\ (1991) and Kano \& Tsuneta (1995, 1996) used it in an abbreviated form, leaving
out the pressure term (i.e., $\beta$~=~0), which does not make a difference in the mathematical treatment, but restricts
the physical applicability. \nocite{kuin-martens82, bray-etal91, kano-tsuneta95, kano-tsuneta96, craig-mcclymont-underwood78, torricelli-ciamponi-einaudi-chiuderi82, landini-monsignori-fossi81}

The energy equation is now cast in a dimensionless form by introducing the variables
\begin{eqnarray}
\eta = (T/T_{max})^{7/2},\\
x=z/L,
\label{eq:dimensionless}
\end{eqnarray}
as well as the parameters
\begin{eqnarray}
\epsilon = \frac{ 2\kappa_0 T^{(11/2+\gamma)}}{7 \chi_{0} P_0^2 L^2}, \\
\xi = \frac{H T_{max}^{2+\gamma +\alpha}}{\chi_{0} P_0^{2-\beta}}, \\
\nu = \frac{2\alpha}{7}, \\
\mu = \frac{-2(2+\gamma)}{7}.
\label{eq:epsilon}
\end{eqnarray}
The result is
\be
\epsilon \eta''\, =\,  \eta^{\mu}\, -\, \xi \eta^{\nu},
\label{eq:eta-de}
\ee
where $''$ represents the second derivative with respect to $x$, while $'$, with
obvious meaning, will also be used below.

Equation~(\ref{eq:eta-de}) is a second order nonlinear ordinary differential equation, to be solved in the domain $x\in[0,1]$ and $\eta \in[0,1]$  There are two parameters $\epsilon$ and $\xi$, that will be shown to be related to the scaling laws.  With the given form of the heating function and the radiative loss function this equation can be solved analytically, as will be done below.

Before doing that I will first discuss the boundary conditions.  It is usually assumed, guided by observations, that the maximum temperature is reached at the top.  Hence the boundary conditions at the loop apex (x=1) become $\eta$(1)=1, and $\eta'$(1)=0.  Since equation~(\ref{eq:eta-de}) is a second order ordinary differential equation these two boundary conditions suffice for determining a solution.  However, the physical nature of the interface between the loop and the underlying chromosphere at the footpoints produces two more boundary conditions.  The temperature of the chromosphere is of the order of 10$^4$ K, while a typical solar coronal loop has an apex temperature of the order of 1-5 MK.  Hence the boundary condition at the bottom satisfies 
$\eta(0)\,=\, (T_{chrom}/T_{max})^{7/2}\, \leq\, 10^{-7}$.  I will show below that for small $\eta(0)$ the
solutions converge towards the solution for $\eta(0) = 0$.  The same is true for the dimensionless conductive flux, $\eta'$.  The vertical extent of the chromosphere is 2000-5000 km, so the average chromospheric  temperature gradient is of the order 3 $\times$ 10$^{-5}$\,K\,cm$^{-1}$.  Hence, for a loop length of the order of 10$^9$ cm, $\eta'(0)\,=\, \frac{(7T(0)^{5/2} dT/dz(0)}{2T_{max}^{(7/2)}/L}\, \approx\, 3 \times 10^{-7}$, and 
again the solution can be developed as a series expansion in that small parameter.  Therefore I can
apply the footpoint boundary conditions $\eta'(0)=\eta(0)= 0$.

These extra boundary conditions imply that usually no solutions exist to equation~(\ref{eq:eta-de}) for arbitrary values of the parameters $\epsilon$ and $\xi$. In general solutions can be found only for one or more discrete values of these parameters\footnote{This is a typical for similar second order differential equations with free parameters, e.g.\ the Schr\"{o}dinger equation, and others of Sturm-Liouville type, such as the Legendre and Bessel equations.}.  I will show below by explicit integration that the extra boundary conditions plus the requirement of a single peaked temperature profile lead to single unique values of $\epsilon$ and
$\xi$.  Writing out these expressions explicitly yields the two scaling laws.  Hence, from a mathematical point of view, the scaling laws are the result of the differential equation for the temperature structure being overdetermined by two excess boundary conditions.  Physically the scaling laws express that for a loop with given length and heating profile, the temperature structure, in particular T$_{max}$ 
and the pressure P$_0$, are fully determined, a result that has been confirmed many times by
numerical simulations (e.g., Serio et al.\ 1981; Aschwanden et al.\ 2001; M\"{u}ller et al.\ 2005).

The solution of equation~(\ref{eq:eta-de}) is straightforward.  Multiplication on both sides with
$\eta'$, and applying the boundary conditions at x=0, results in the first integral,
\be
\frac{\epsilon\eta'^2}{2}\, =\, \frac{\eta^{\mu+1}}{\mu+1} - \frac{\xi \eta^{\nu+1}}{\nu+1}.
\label{eq:order1-de}
\ee

Applying the remaining boundary conditions at x=1 produces a generalized scaling law, usually
called the second scaling law,
\be
\xi\, =\,  \frac{\nu+1}{\mu+1}\, =\, \frac{2\alpha+7}{3-2\gamma},
\label{eq:ksi-scaling}
\ee
which shows that $\xi$ takes on a specific numerical value for a given heating mechanism ($\alpha$)
and fit to the radiative loss function ($\gamma$).
I will demonstrate below that this indeed reproduces the second scaling law in the more familiar form.  Inserting the formal boundary conditions $\eta(0)$ and  $\eta'(0)$ introduces only vanishing corrections to this scaling law
\be
\frac{\epsilon(\eta'^2 - \eta'(0)^2)}{2}  =  \frac{\eta^{\mu+1}}{\mu+1} - 
\frac{\eta(0)^{\mu+1}}{\mu+1}
- \frac{\xi\eta^{\nu+1}}{\nu+1} + \frac{\xi\eta(0)^{\nu+1}}{\nu+1}.
 \ee

Applying the boundary conditions at x=1 the second generalized scaling law for arbitrary conditions at the
foot points of the loop is found
\be
\xi = \frac{\nu+1}{\mu+1}\, \frac{1-\eta(0)^{\mu+1}+\epsilon \eta'(0)^2 (\mu+1)/2}{1-\eta_0^{\nu+1}}.
\label{eq:scaling1}
\ee
For $\gamma$=0.5, $\mu$ =-5/7, and for $\alpha\, \geq$ -7/2, $\nu$+1 $\geq$ 0.  Hence there are no
singularities in equation~({\ref{eq:scaling1}) under these conditions, and the correction terms are vanishingly
small for realistic chromospheric boundary conditions.

What remains is the first order nonlinear ordinary differential equation~(\ref{eq:order1-de}).  It can be expressed as
\be
\eta'\, =\, [\frac{2(\eta^{\mu+1}-\eta^{\nu+1})}{\epsilon(\mu+1)}]^{1/2},
\ee
where the positive root has been taken conform with a monotonic rise in $\eta$ (end hence temperature)
from the bottom to the top of the loop.  This equation can be directly integrated and yields
\be
x\, =\, [\epsilon(\mu+1)/2]^{1/2}\, \int_{0}^{\eta}\frac{dy}{(y^{\mu+1}-y^{\nu+1})^{1/2}}.
\label{eq:integral}
\ee
This expression can be simplified by the transformation
\be
u\, =\, \eta^{\nu-\mu},
\label{eq:def-u}
\ee
with a similar substitution of $y$ for $z$ in the integrand.  The integral now reduces to
\be
x\, =\, \frac{[\epsilon(\mu+1)/2]^{1/2}}{\nu-\mu} \int_{0}^{u}\frac{z^{\lambda}dz}{(1-z)^{1/2}},
\label{eq:integrand}
\ee
where the parameter $\lambda$ is defined as
\be
\lambda\, =\, \frac{1-2\nu+\mu}{2(\nu-\mu)}.  
\label{eq:lambda1}
\ee
The integrand in equation~(\ref{eq:integrand}) represents the definition of the well-known incomplete
$\beta$-function (e.g.,  Zwillinger,1996, p.\ 497),  i.e.,  \nocite{zwillinger96}
\be
\int_{0}^{u}\frac{z^{\lambda}dz}{(1-z)^{1/2}}\, =\, \beta(u;\lambda+1,1/2),
\ee
where Re($\lambda$) $>$ -1, so that the integral does not diverge.  In the appendix it will be shown 
that the effect of a non-zero chromospheric temperature at the lower boundary is very minor.
Restoring the original parameters $\alpha$ and $\gamma$, one finds
\begin{eqnarray}
\lambda\, =\, \frac{(3/2)-\gamma-2\alpha}{2(2+\gamma+\alpha)},\\
x\, =\, [\frac{\epsilon(1-2(2+\gamma)/7)}{2}]^{1/2}\, \frac{7\beta(u;\lambda+1,1/2)}{2(2+\gamma+\alpha)}.
\label{eq:x-run}
\end{eqnarray}
Thus the condition $\lambda$ $>$ -1 implies $\alpha$ $>$ -(2+$\gamma)$, i.e., $\alpha$ $>$ -2.5 for
$\gamma$=0.5.

The boundary conditions at the bottom of the loop are satisfied by construction, and applying the
boundary conditions at the top of the loop yields
\be
1\, = \, [\frac{\epsilon(1-2(2+\gamma)/7)}{2}]^{1/2}\, \frac{7B(\lambda+1,1/2)}{2(2+\gamma+\alpha)},
\ee
where B is the complete $\beta$-function (i.e., the integral carried out up to u=1).
This condition sets the value of the parameter $\epsilon$, which, as I shall show below, defines
the first (RTV) scaling law,
\be
\epsilon\, =\, \frac{2}{(1-2(2+\gamma)/7)}\, [\frac{2(2+\gamma+\alpha)}{7}]^2\, \frac{1}{B^2(\lambda+1,1/2)}.
\label{eq:epsilon-scaling}
\ee
Eliminating $\epsilon$ from equation~(\ref{eq:x-run}) yields
\be
x\, =\, \beta_r(u;\lambda+1,1/2),
\ee
where $\beta_r$ is the regularized incomplete $\beta$-function (i.e., divided by the corresponding
complete value).

Inverting this equation and restoring the original dimensional variables produces an explicit 
expression for all temperature profiles defined by valid choices of the parameters $\alpha$ and
$\beta$ in the heating function $E_h=HP_0^{\beta}T^{\alpha}$,
\be
T(z)\, =\, T_{max}[\beta_r^{-1}(s/L;\lambda+1,1/2)]^{\frac{1}{2+\gamma+\alpha}},
\label{eq:t-profile}
\ee
where $\beta_r^{-1}$ is the inverse of the regularized incomplete $\beta$-function.  
Note that the parameters H, $P_0$, and $\beta$ from the heating function do not appear in this
solution.  I will show below that H and $P_0$ do show up in the scaling law relations for $T_{max}$,
and hence that they do influence the solution.

\subsection{Solutions with Temperature Reversals \label{sec:nonuniform}}

When the condition of a uniform increase in temperature is dropped but the condition of $\eta'$~=~ 0
at the boundaries retained, an infinity of solutions is available.  The second possible solution is one with
a temperature maximum at x~=~1/2, and a vanishing temperature and conductive flux
($\eta$~=~$\eta'$~=~0) at x~=~1.   The third solution is one with a temperature maximum at x~=~1/3,
a minimum at x~=~2/3, and again a maximum at x~=~1, and solutions for higher eigenvalues are in the
same vein.  The segments between the temperature minimum and maximum are each still the inverse
of the regularized $\beta$-function, with x~$\rightarrow$~-x for the segments with decreasing temperature.

This series of solutions with an infinite number of eigenvalues is typical for the type of second order
ordinary differential equations that is being considered here, and the analogous sets of solutions
for the Schr\"{o}dinger, Bessel, and Legendre equations are well known.  Note that in contrast to the
eigenfunctions of those examples, the inverted regularized $\beta$-functions are not orthogonal,
since they are non-negative.

With $n$ denoting the order of the solution, the expression for $x$ of the first segment (eq.~[\ref{eq:x-run}])
becomes
\be
x\, =\, [\frac{\epsilon_n(1-2(2+\gamma)/7)}{2}]^{1/2}\, \frac{7\beta(u;\lambda,1/2)}{2(2+\gamma+\alpha)},
\label{eq:epsilon-multiple}
\ee
where $\epsilon_n$ is the value that $\epsilon$ must take to make this solution possible (i.e.,
the eigenvalue).  Applying the boundary condition at the end of the first segment ($x\,=\, 1/n$) yields
the expression for $\epsilon_n$  (crf.\ eq.~[\ref{eq:epsilon-scaling}]).
\be
\epsilon_n\, =\, \frac{2}{n^2(1-2(2+\gamma)/7)}\, [\frac{2(2+\gamma+\alpha)}{7}]^2\, \frac{1}{B^2(\lambda+1,1/2)}.
\label{eq:epsilon-n-scaling}
\ee

The physical interpretation of this result will be discussed in \S\,~\ref{sec:multiple-t}.

\section{Physical Interpretation \label{sec:interpretation}}

Having developed a simple analytical expression for the temperature profile of an elementary coronal 
strand for a whole array of heating mechanisms, I will now consider the physical interpretation of this
result.  First I will derive the scaling laws in their better known dimensional form.  Then I will explore the 
parameter space of the solutions for different heating functions, and discuss how loop observations may 
serve as a diagnostic for determining the heating function.  The specific example of Joule heating will be
treated in detail, followed by an analysis of three other coronal heating mechanisms that are captured
by the formalism of this paper.  Finally I will derive a solution in which the assumption of constant loop
cross-section is relaxed and show that only marginally affects earlier conclusions.

\subsection{Scaling Laws}

As was pointed out in the previous section, the differential equation for the temperature profile of a 
coronal strand is over-constrained by its four physical boundary conditions.  The consequence is that 
solutions only  exist for a specific value of each of the two parameters ($\epsilon$ and $\xi$), that was 
determined above.  Directly therefrom the scaling laws are derived.  Writing out the relation
for $\epsilon$, equation~(\ref{eq:epsilon-scaling}), yields
\be
P_{0}L\, =\, T_{max}^{\frac{11+2\gamma}{4}}\, [\frac{\kappa_{0}}{\chi_{0}}]^{1/2}\, 
\frac{(3-2\gamma)^{1/2}}{4+2\gamma+2\alpha}\, B(\lambda+1,1/2).
\label{eq:first-scaling}
\ee
Note that is very close to the original RTV scaling law for uniform heating.  For $\gamma$ = 1/2 one finds
$P_{0}L\, \thicksim\, T^3$, the original result.  The only place where the functional form of the heating
function plays a role is in the constant of proportionality, but  there it will be found to be of great  potential 
significance as a diagnostic.   Craig et al.\ (1978), and later Kuin \& Martens (1982), derived that the form of
the scaling laws is preserved for the non-uniform heating considered here, and Bray et al.\  (1991, p.\ 297)
calculated the constant of proportionality explicitly.  Kano \& Tsuneta
(1995, 1996) included a non-zero conductive flux at the loop apex in order to model looptop heating -- as one
would expect in flares -- therefrom deriving an upper and lower limit to the constant of proportionality in the
first scaling law.

The expression for $\lambda$ is given in equation~(\ref{eq:lambda1}).  Expressing this relation in terms
of the parameters in the heating function yields
\be
\lambda+1\, =\, \frac{11/2+\gamma}{2(2+\gamma+\alpha)}.
\ee
Thus for $\gamma$=1/2, $\lambda$+1=6/(5+2$\alpha$), which is a decreasing function of $\alpha$.
The first scaling law, equation~(\ref{eq:first-scaling}), demonstrates that for a loop with a given length
and maximum temperature the loop pressure, P$_0$ is proportional to 
$B(\lambda+1,1/2)/(4+2\gamma+2\alpha)$, and Figure~\ref{fig:over-pressure} shows the
degree of over-pressure for loops with heating concentrated at  the footpoints ($\alpha< $0),
and v.v.\ the under-pressure of strands heated at the apex.  Note that over-pressure with
respect to the uniform heating is therefore consistent with the scaling laws, and indeed
quantified by them.  Over-pressure in loops with footpoint heating is a feature that has been
reported frequently from hydrodynamical simulations of coronal loops (e.g., Serio et al.\ 1981;
Aschwanden et al.\ 2001; M\"{u}ller, Hansteen, \& Peter 2003;  M\"{u}ller, Peter, \& Hansteen 2004;
M\"{u}ller et al.\ 2005).

\begin{figure}[!h]
\epsscale{1.0}
\plotone{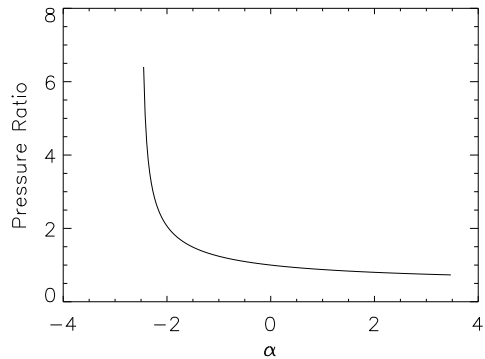}
\caption{Over-pressure with respect to the RTV scaling law for uniform loop heating 
as a function of the parameter $\alpha$ in the non-uniform heating function of this paper, $E_h = H P_0^{\beta} T^{\alpha}$.
\label{fig:over-pressure}}
\end{figure}

I further note that the measurement of over-pressure in elementary strands (as defined in the 
introduction), is, in principle, a diagnostic for the form of the heating profile, at least for
negative  $\alpha$'s, since it 
directly leads to an estimate of $\alpha$.  However, when a bundle of elementary strands fills 
an observed loop with an unknown filling factor, the emission measure of the loop cannot,
by itself, lead to a  correct estimate for the density, and hence the pressure (e.g., Martens,
van den Oord, \& Hoyng 1985, Porter \& Klimchuk 1995).
\nocite{martens-vandenoord-hoyng85, porter-klimchuk95}
Spectral information from density sensitive line ratios may be the solution here, but then
there still may be confusion from 
the line-of-sight addition of several strands.   Below I will show that the expansion of the loops 
from footpoint to top (cfr.\ DeForest, 2007) \nocite{deforest07} does not influence the
over-pressure in a significant way, and hence strand pressure is a good diagnostic for the 
heating function.  A word of caution must be added here though, and that is that this result only 
applies to quasi-static strands.  Warren, Winebarger, \& Hamilton (2002) and Winebarger, 
Warren, \& Seaton (2003) have demonstrated that strands that
\nocite{warren-winebarger-hamilton02, winebarger-warren-seaton03}
are cooling down, e.g., in the aftermath of a sudden heating event like a flare, tend to have
over-pressures with respect to the scaling law for uniform heating as well, and by a far greater
degree than footpoint heated loops

The second scaling law is derived by writing down the relation for $\xi$, equation~(\ref{eq:ksi-scaling}), 
explicitly.  In terms of  the original parameters for the heating function the result is
\be
\frac{HT_{max}^{2+\gamma+\alpha}}{\chi_0P_0^{2-\beta}}\, =\, \frac{7/2+\alpha}{3/2-\gamma}.
\label{eq:second-scaling}
\ee
For uniform heating, $\alpha$=$\beta$=0, one finds back the RTV second scaling law, 
$H=E_h\, \thicksim\, P^{7/6}/L^{5/6}$, after elimination of $T_{max}$ with the use of the first
scaling law.  A more intuitive way of expressing the second scaling law above is in terms of the 
heating rate at the top of the strand,
\be
E_{top}\, =\, HT_{max}^{\alpha}P_0^{\beta}\, =\, \frac{P_0^2\chi_0(7/2+\alpha)}
{T_{max}^{2+\gamma}(3/2-\gamma)}.
\label{eq:e-top}
\ee
The two scaling laws completely define the quasi-static hydrodynamic solution for the strand,
i.e., one can algebraically derive the loop pressure and temperature profile from the heating 
function -- the apex heating rate plus parameters $\alpha$ and $\beta$ -- and the loop length.

Finally, instead of normalizing the heating function by the apex heating, one might want to 
express this function in terms of the Poynting flux $F$ through the footpoints.  This is 
accomplished as follows
\be
F\, =\, \int_0^L E_h(s)ds\, =\, H L P_0^{\beta} T_{max}^{\alpha} \int_{0}^1 t^\alpha dx.
\label{eq:flux-integral}
\ee
Replacing $dx$ using the differential form of equation~(\ref{eq:integrand}) yields an integral that
can be transformed into a $\beta$ function in the same manner as in
Section~\ref{sec:formalism}.  The final result is
\be
F\, =\, \int_0^L E_h(s)ds\, =\, H L P_0^{\beta} T_{max}^{\alpha} \frac{B(\sigma+1,1/2)}{B(\lambda+1,1/2)},
\label{eq:flux}
\ee
where the parameter $\sigma$ is defined as
\be
\sigma\, =\, \lambda + \frac{\alpha}{\alpha+\gamma+2}.
\ee
The requirement $\sigma$ $>$ -1 translates into $\alpha$ $>$ -(2+$\gamma$), which is the
same requirement as for $\lambda$.  

Equation~(\ref{eq:flux}) above can be used to eliminate the constant $H$ in the heating function in favor
of the, in many cases, physically more meaningful Poynting flux entering from the chromosphere. 
Kano \& Tsuneta (1995) have derived  such an expression without solving for the temperature profile.
After eliminating H the
expressions for the hydrodynamical variables T$_{max}$ and P$_0$ in terms of the magnetic
field parameters loop-length and Poynting flux, as well as heating function parameters $\alpha$
and $\beta$ become
\be
T_{max}\, =\, (\frac{FL}{\kappa_0})^{2/7} 
[\frac{4(2+\gamma)^2}{(7/2+\alpha)B_{\lambda}B_{\sigma}}]^{2/7},
\label{eq:tmax}
\ee
and
\begin{eqnarray}
P_0\, =\, \frac{F^{(11+2\gamma)/14} L^{(2\gamma-3)/14}}{\kappa_0^{(2+\gamma)/7}\chi_0^{1/2}} \times \\
            \frac{2(2+\gamma+\alpha)}{(3-2\gamma)^{1/2}B_{\lambda}}^{2(2+\gamma)/7}
            [\frac{(3-2\gamma)B_{\lambda}}{(7+2\alpha)B_{\sigma}}]^{(11+2\gamma)/14}, \nonumber
\label{eq:p0}
\end{eqnarray}
where the meaning of the shorthand B$_{\lambda}$ and B$_{\sigma}$ is obvious
(cfr.\ eq.~[\ref{eq:flux}]).

Note that only the expressions between square brackets depend on the details of the heating
distribution, while the parameter $\beta$ from the heating function does not appear at all.  One
interesting inference from this result is that the loop pressure is almost proportional to the
Poynting flux entering into and dissipated in the loop, but only very weakly dependent on the loop
length ($\thicksim$ L$^{-1/7}$).  The maximum temperature of the loop is not strongly
dependent on either:  for a loop with a given length an increase of incoming flux of an order of
magnitude only results in slightly less than a doubling of the temperature.  Hence T$_{max}$ is
a relatively insensitive diagnostic for the incoming flux, unlike the loop pressure.  The fact that
there is a large spread in observed loop apex temperatures, from about 1 to 5 MK, then implies
that there is an enormous range in incoming Poynting flux for solar coronal loops.

\subsection{Temperature Profiles as a Heating Diagnostic}

One might expect that vastly different heating functions (e.g., loop apex heating versus footpoint
heating) would give rise to temperature profiles that are sufficiently different from each other to 
allow for the determination of the heating profile from a series of temperature measurements 
along the axis of an elementary strand.  This turns out not to be the case as is demonstrated in 
Figure~\ref{fig:t-profile}.  The left hand panel of this figure shows the run of the dimensionless 
temperature for ten values of $\alpha$, defined by
\be
\alpha(n)\, = -2.0 + n/2;\ n = 0, 1,......,9.
\label{eq:alphas}
\ee
It is immediately obvious that the differences between the temperature profiles are very small, 
and that the inversion (i.e., reproducing the heating function from the temperature profile) would 
be very hard to do with real data.  The flattest profiles (relatively) are those for heating concentrated
at the footpoints, but their flatness is never equal to that derived in the numerical solutions of 
Aschwanden et al.\ (2001).  An investigation into the reason for this discrepancy
will be reported in a future paper.

\begin{figure*}[!t]
\epsscale{2.00}
\plottwo{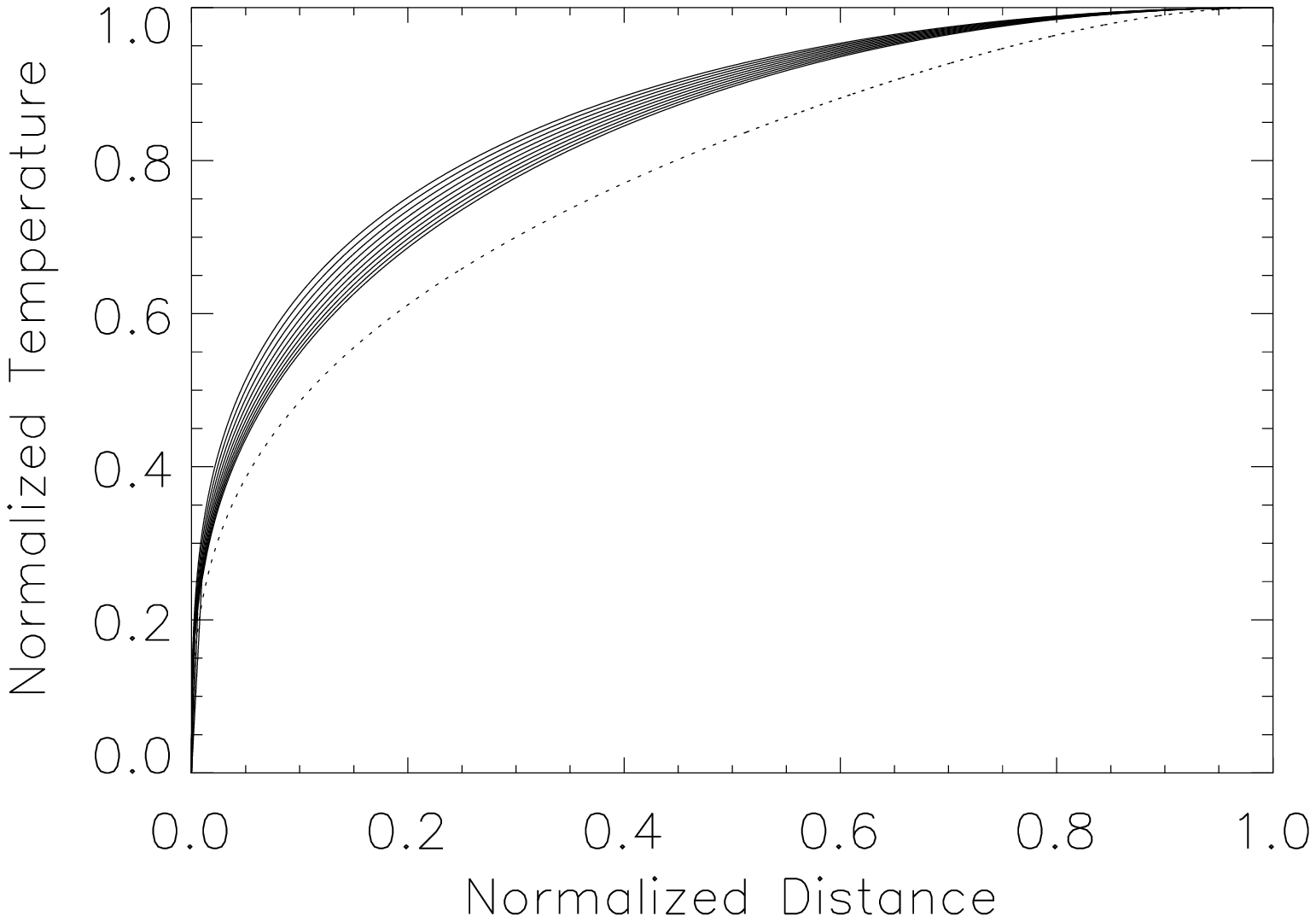}{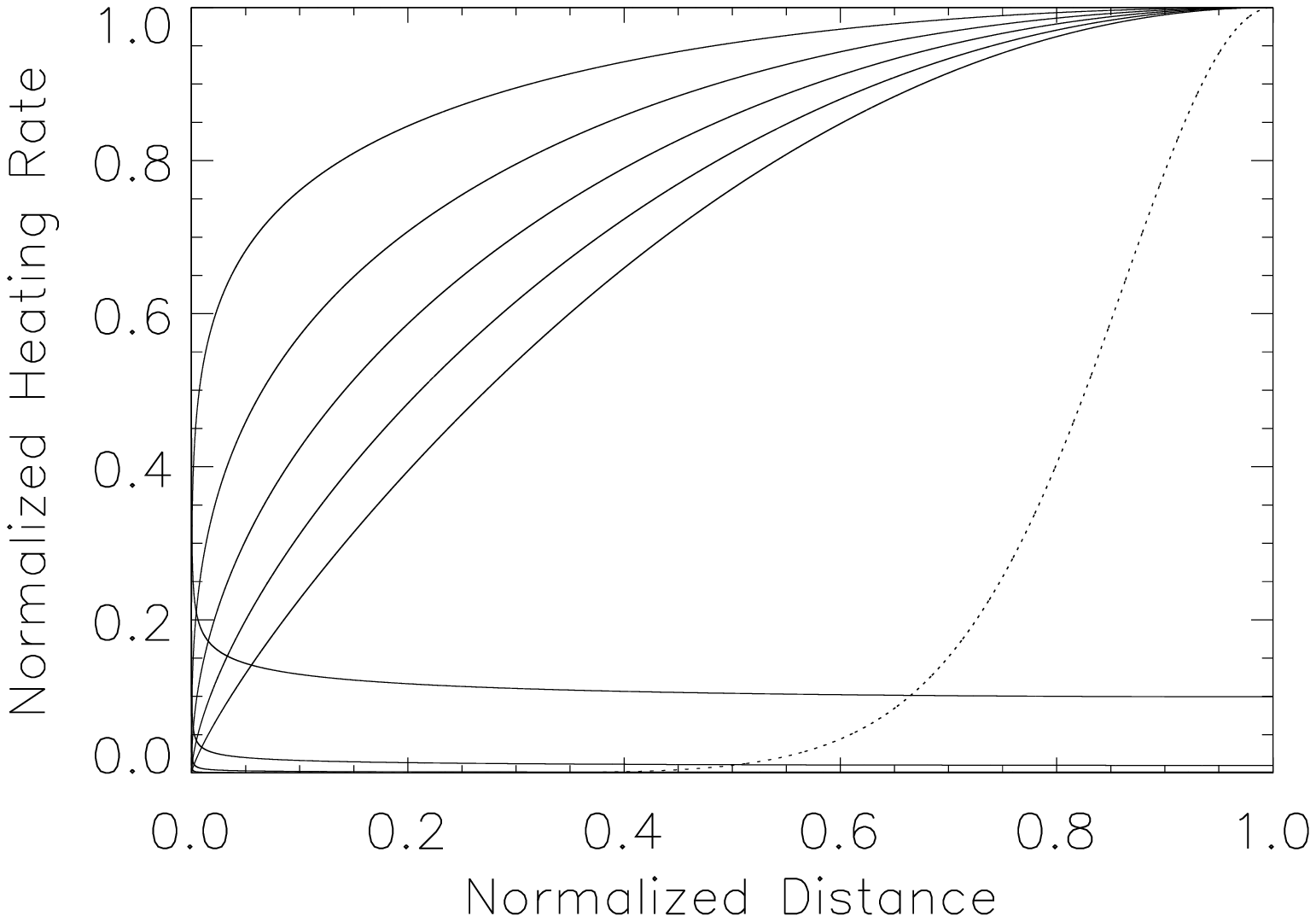}
\caption{Left:  Dimensionless temperature solutions for the non-uniform heating function
$E_h = H P_0^{\beta} T^{\alpha}$, with parameters from equation~(\ref{eq:alphas}).
The dashed line corresponds to $\alpha$ = 25.  The lowest value of $\alpha$ corresponds
to the flattest curve (i.e., the upper one at t=0.5), and the temperature profiles become less flat as
$\alpha$ increases.  Right:  The dimensionless heating functions corresponding to the temperature
profiles on the  left.  For negative values of $\alpha$ the functions are normalized at t=0.01, and for
positive values of $\alpha$ at the apex. The lowest value of $\alpha$ corresponds to the largest ratio
of footpoint to apex heating, with the ratio monotonically decreasing for increasing value of $\alpha$. 
The dashed line represent the heating profile for $\alpha$ = 25.
\label{fig:t-profile}}
\end{figure*}

The right hand panel of Figure~\ref{fig:t-profile} shows the heating functions for the same values of
$\alpha$ as in the left hand panel and the equation above.  The heating functions for $\alpha\,>$\,0 are
normalized by their apex value, while for $\alpha\,<$\,0 the heating is normalized at a footpoint
temperature  t$_0$\,=\,0.01, i.e., 1\% of the maximum temperature at the apex.  The panel shows that the
heating becomes more and more concentrated near the footpoints as $\alpha$ becomes more negative.
There is no analytical solution for $\alpha$=-2.5 or less, and the stability of the solutions with $\alpha\,
\leq$ -2 is doubtful.  The results of a full numerical stability analysis are currently under investigation and
will also be reported in a future paper.  The solution for uniform heating is stable and within a few percent
of the numerical solution for the full radiative loss function and including gravity, as reported in
Martens et al.\ (2000).

Hence, the conclusion of this section is that, in contrast to the scaling laws, the temperature profiles of 
elementary strands are of limited value for the determination of the heating profile.  The exception to
that might be the case where the heating is strongly concentrated near the top of the loop.  This situation
is captured in a basic sense in the solutions presented by Kano \& Tsuneta (1995, 1996)
\nocite{kano-tsuneta95, kano-tsuneta96}
who chose a non-zero conductive flux at the top of the loop in order to simulate a large heat influx
from the top, a situation that may apply in the case of flaring loops.  Priest et al.\ (1998, 2000) present
approximate solutions for the temperature profile for different heating profiles and claim
therefrom that apex heating can be discerned from uniform and footpoint heating from the temperature
profile.  Mackay et al.\ (2000) dispute this by demonstrating that the errors in the measurements are too
large to make a meaningful distinction.
\nocite{priest-etal98, priest-etal00, mackay-etal00b}

Here I have simulated heating concentrated around the loop apex by using a large value for the heating
parameter $\alpha$.  The dashed line in Figure~\ref{fig:t-profile} represents the solution for the case
$\alpha$~=~25.  It clearly exhibits a peak in the heating profile around the temperature maximum, but
the temperature profile is only slightly different from the other solutions presented in the figure.  I
conclude therefrom that in principle one might indeed distinguish between apex heating from uniform
and footpoint heating using the temperature profile, but that in practice the measurement errors will
likely interfere, which agrees with both Priest et al.\ and Mackay et al.
However, heating concentrated near the loop apex may apply to flaring loops, but, as will be shown
below, typical heating mechanisms for non-flaring Active Region loops do not have large values of the
heating parameter $\alpha$.

\subsection{Joule Heating}

As a specific example of  a physical heating mechanism that is captured by the formalism developed
above I will now consider Ohmic dissipation of DC coronal currents. 
Joule heating is described by
\be
E_h\, =\, \eta_S j^2\, 
\label{eq:joule-heating}
\ee
where $j$ is the coronal current density, and $\eta_S$ the Spitzer resistivity.  (The subscripts ``S" 
and ``0" are used here to avoid confusion with the variable $\eta$ used throughout this paper).
The Spitzer resistivity depends on the local hydrodynamical variables as (Spitzer 1962; 
Allen 1973, p.\ 50)
\nocite{allen73, spitzer62}
\be
\eta_S\, =\, \eta_0 \ln(\Lambda) T^{-3/2}
\label{eq:spitzer}
\ee
with $\Lambda$ denoting the plasma parameter, and  $\eta_0$ a constant.
Although $\Lambda$ (the number of particles in a Debye sphere) depends on the local density and
temperature (e.g., Sturrock 1994, p.\ 11), \nocite{sturrock94} I will ignore that dependence since it
matters little after taking the natural logarithm.  For typical coronal conditions, T = 2  MK and density
n = 3 $\times$ 10$^9$ cm$^{-3}$, I find $ln(\Lambda)$ = 16.7.  $\eta_0$ is equal to $9.0\times10^{-9}$
esu (Allen 1973, p.\ 50).
\nocite{allen73}

In terms of the generalized heating function used above one finds $\alpha$ = -1.5, $\beta$ = 0,  and
H = $\eta_0 j^2 ln(\Lambda)$.  In a force-free coronal flux tube with constant cross-section the current
density $j$ is indeed constant along the tube, so H is nearly constant.
The solution for the temperature profile and the heating function is shown by one of the curves
in Figure~\ref{fig:t-profile} (the second flattest temperature profile, and the second-most 
peaked at the footpoints heating function).  The amount of over-pressure in a Joule heated
coronal strand is a factor of about 1.5 according to Figure~\ref{fig:over-pressure}.  That results in an
excess brightness of a factor 2.2 compared to a uniformly heated strand.

The heating rate in coronal loops that is inferred from the observed radiation and the scaling laws is
of the order of 10$^{-2-3}$ erg\,cm$^{-3}$\,s$^{-1}$.  Using the value of $\eta_0$ given above and a
temperature of 2 MK, I find that a current density of 7.1\,$\times$\,10$^6$ statamp cm$^{-2}$ is
required.  That corresponds to an electron drift velocity of about 50 km/sec, which is indeed much 
less than the coronal electron thermal velocity, the condition for Spitzer resistivity to apply when
the ion temperature equals the electron temperature.  Note that down the
legs of the loop the ratio of the drift velocity to the electron thermal velocity decreases further,
since the drift velocity scales as T in a strand with constant pressure and current density, while the
electron thermal velocity scales only as $\sqrt{T}$.  This implies that Spitzer heating will reach its
limit first at the loop apex, so that anomalous resistivity, leading to further enhanced heating, will start
at that location and expand throughout the corona before it penetrates the chromosphere.   

It needs to be pointed out that the current density derived above is much higher than the current
density inferred from vector magnetograph observations.  Pevtsov et al.\  (1994, 1995, and private
communications) report an average unsigned force-free $\alpha$ per pixel  of the order of
10$^{-9}$ cm$^{-1}$ in active regions.  This is an order of magnitude larger than their average $\alpha$
in active regions, because $\alpha$'s of
both signs are present.  Each active region appears to be subdivided in a handful of patches with same
sign $\alpha$  (Pevtsov, private communication). Since the force-free $\alpha$ is constant along a
field line, it has the same value in the corona.  With an extrapolated coronal field strength of the order of
100 Gauss one finds from Amp\`{e}re's law $j$ = 240 statamp cm$^{-2}$, more than four orders of
magnitude shy of the value required for Ohmic heating derived above.
\nocite{pevtsov-canfield-metcalf94, pevtsov-canfield-metcalf95}

However, the measurements of Pevtsov et al.\  at the Haleakela Stokes Polarimeter are samples with
a pinhole of approximately a six arcsecond diameter, and the currents derived therefrom are totals over
the whole pixel, not really current densities.   Hence it is possible, in principle, that there are sheets with
very high current densities in each pixel, with nearly current-free regions in between, that produce the
required heating rate, although at the obvious expense of a reduced filling factor.
It is easily verified that in a 100 Gauss field a single sheet with a thickness of the order of 340 m and a
current density of 7.1\,$\times$\,10$^6$\, statamp cm$^{-2}$, as derived above, would already produce
a current density of twice the 240 statamp cm$^{-2}$ per pixel implied from the Pevtsov et al.\ results.  
The filling factor can be enhanced by introducing a larger number of sheets with counter-flowing currents.
The work of van Ballegooijen (1986) on current
filamentation resulting from regular footpoint photospheric motions, and the work by Longcope (1996)
on current
generation on magnetic separators, both strongly suggest that coronal currents will be highly non-uniform.
\nocite{vanballe86, longcope96}  Hence coronal heating by Ohmic dissipation in a number of unresolved
high current density sheets inside coronal loops is a candidate heating mechanism that cannot be dismissed
out of hand.

\subsection{Anomalous Resistivity Phase-mixing, Nano-flares, and Velocity Filtration  \label{sec:other-heating}}

It was mentioned above that a number of coronal heating mechanisms proposed in the literature can be
captured by the heating parametrization used in this paper, with the heating rate proportional to density
(or pressure) and temperature, each to an arbitrary power (eq.~[\ref{eq:heatingf}]).  In this section I will
briefly describe a number of these in addition to regular Joule heating with Spitzer resistivity discussed
above.  This discussion does not constitute a comprehensive review.

First let's consider anomalous current dissipation driven by the ion-acoustic instability, which occurs when
the electron drift velocity surpasses
the electron thermal velocity of the plasma.  In that case the resistivity is anomalously enhanced
over the Spitzer resistivity and the current dissipation and heating rate can greatly increase.  Rosner et al.\
(1978a)  \nocite{rosner-etal78} discuss this scenario in depth and cite a resistivity of the order of
\be
\eta\, =\, 2\times 10^{-8} n_{e}^{-1/2} s
\label{eq:ion-acoustic}
\ee
where $n_{e}$ is the electron density in cgs units.  As shown above the current density $j$ is constant
along the strand and hence, using the gas law, the heating rate $E_{h}\, =\, \eta j^2$, scales as
$T^{1/2} P^{-1/2}$.   Therefore, in the formalism of this paper $\alpha$ = 1/2 and $\beta$ = -1/2 for
ion-acoustic heating.

Viscous dissipation of shear Alfv\'{e}n waves is a viable heating mechanism proposed by Heyvaerts \&
Priest (1983). \nocite{heyvaerts-priest83}  The physical basis is that standing Alfv\'{e}n waves on neighboring
field lines will rapidly grow out of phase even for a slightly varying Alfv\'{e}n speed across the magnetic field,
a process called phase-mixing.  This produces large velocity gradients that result in viscous dissipation
of the wave velocity fields.  The heating rate in the Heyvaerts and Priest mechanism is given by
\be
E_h\, =\, \mu_{v} \rho v_{wave}^2/d^2
\label{eq:phase-mixing}
\ee
with $\mu_{v}$ representing the shear viscosity, $\rho$ the density, $v_{wave}$ the velocity amplitude of the
 Alfv\'{e}n wave, and $d$ the typical cross-field length-scale over which the wave velocity varies.  Heyvaerts
and Priest demonstrate that the cross-field length-scale $d$ decreases during the process of phase-mixing
until the total wave dissipation rate balances the incoming Poynting flux from the Alfv\'{e}n waves.

It may be assumed that the kinetic energy density in the standing Alfv\'{e}n waves is constant along the
loop, i.e., $ \rho v_{wave}^2$ = constant.  There is indeed some observational support for this in that
the measured square of the non-thermal broadening velocity in Active Region emission lines scales
linearly with temperature (Saba \& Strong 1991), \nocite{saba-strong91} although these data from
the Flat Crystal Spectrometer onboard the {\it Solar Maximum Mission} ({\it SMM})
have very low resolution ($\thicksim$~15 arcsec) so that individual loops are not resolved.   Combining the
gas-law with energy equipartition in the wave ($ \rho v_{wave}^2$ = constant) reproduces this scaling. 
Using the constancy of $d$ and $ \rho v_{wave}^2$ throughout
the loop I find that the heating scales with the shear viscosity $\mu$.  Since  $\mu$ depends on plasma
density and temperature as $T^{5/2}$ (e.g., Choudhuri 1998), the heating rate from the phase-mixing
model is characterized by $\alpha$ = 5/2 and $\beta$ = 0.  \nocite{choudhuri98}

Note that according to equation~(\ref{eq:first-scaling}) and Figure \ref{fig:over-pressure}, for a loop strand with
a given (observed) length and maximum temperature the gas pressure for the case of Joule heating
exceeds that for the case of heating via phase-mixing by a factor two.  Hence a measurement of strand
pressure, e.g.\ by taking measurements of density sensitive line pairs at a given temperature, is a
possible diagnostic for discriminating between these two heating mechanisms, once the elementary
strands have been resolved.

Uniform heating, i.e., $\alpha$ = 0 and $\beta$ = 0, is the type of heating often considered in numerical
simulations, and also in the original scaling-law papers by Rosner et al.\ (1978b) and Craig et al.\ (1978). 
The reason is of course that
this is the simplest assumption, but also that the widely known theory of nano-flare heating of the solar
corona (e.g., Parker 1988, 2005) is compatible with uniform heating. \nocite{parker88, parker05}
Because of the high plasma $\beta$ in coronal loops the origin and development of nano-flares is
considered by Parker as a purely magnetic phenomenon, i.e., not dependent in first order upon local
thermodynamic quantities such as density and temperature.  Hence it is reasonable to assume that the
distribution of the nano-flares is uniform throughout loop strands.  Moreover, in order to reproduce the observed
quasistatic appearance of Active Region loops (Antiochos et al.\  2003), for each loop strand the time
between nano-flares has to be shorter than thermal cooling time of the strand.  And indeed, because the loop
thermal cooling time acts as an effective frequency filter in the hydrodynamical evolution of loops, nano-flares
with a frequency that is sufficiently smaller than the frequency associated with the cooling time, will appear, to
first order, as steady heating.  Of course there will be effects such as Doppler-shifts and line-broadening due
to the explosive nature of the nano-flares., which can be used as diagnostics to distinguish between
nano-flare heating and true steady heating (e.g., Patsourakis \& Klimchuk, 2006) but the thermal structure
of the strands will be identical to first order.  \nocite{patsourakos-klimchuk06}

Finally, some heating mechanism are considered non-local, in the sense that free magnetic energy is converted
into particle kinetic energy at a different location from where particle kinetic energy is thermalized in the
background plasma.  Thus the conversion of magnetic to particle energy takes place in a different location
than the heating of the background plasma, and hence the term non-local, although  the heating itself still
occurs in-situ.   One example of this is the heating
of flaring loops resulting from the impact of electron beams that are often assumed to have been
accelerated in a current sheet on top of these loops.   A second example, of potential relevance to the heating
of quasi-steady Active Region coronal loops, is the filtration of the high energy non-thermal tails in
the electron distribution from the transition region into the overlying corona (Scudder 1992a, 1992b, 1994;
Vi\~{n}as, Wong, \& Klimas 2000). \nocite{scudder92a, scudder92b, scudder94, vinas-wong-klimas00}
We then have a situation in which high energy electron beams enter coronal loops through their footpoints,
emanating, paradoxically, from the much cooler transition region and chromosphere.   Scudder points out
that this effect is possible because of  the much higher particle density in the chromosphere and transition
region.  Since the mean-free-path of the electrons in the filtration beams scales with the fourth power of
the beam electron velocity, it only takes a relatively small ratio of beam velocity to thermal electron velocity
in the corona for the mean-free-path of the beam electrons to surpass the loop length and hence for the
filtration beams to travel through the loop strands relatively unattenuated.   In that case the number of 
collisions between beam and background plasma particles at any location in the loop is proportional to
the density alone,  and therefore the heating rate is as well.  In the formalism of this paper that results in
the parametrization $\alpha$ =  -1 and $\beta$ = 1.

In closing this section I note that the coronal heating mechanisms discussed here (anomalous current
dissipation, phase-mixing, nano-flares, and velocity filtration) are all relatively old.  The reason is, I believe,
that after it became clear that coronal heating through the dissipation of acoustic waves was implausible
(Athay \& White 1979), \nocite{athay-white79}
solar theorists picked the low-hanging fruits first, i.e., the most obvious and
plausible heating mechanisms were investigated first.  More current papers on coronal heating often
continue the development of older ideas, some of which are described here, or pursue more exotic
possibilities.  Solar physics is now faced with the situation of a plethora of possible coronal heating
mechanisms, which, as the current paper illustrates, are difficult to distinguish observationally since
the temperature profiles are almost identical, and the predicted loop pressures vary by amounts that are
currently difficult to derive from measurements of unresolved multi-stranded loops.  However, given
this multitude of theoretical possibilities, I disagree that our inability to pinpoint the actual coronal
heating mechanism forces the conclusion that there is no in-situ coronal heating at all
(Aschwanden et al.\ 2007)! \nocite{aschwanden-etal07}

\subsection{Loops with Variable Cross-Section \label{sec:variable}}

As was pointed out in the introduction, a recent paper by DeForest (2007) \nocite{deforest07} has made
a powerful argument that the elementary strands of at least EUV emitting loops expand with distance along
the loop axis, as one would expect based on simple potential or force-free magnetic field extrapolations.
The formalism developed in this paper can be used to capture loop expansion with height in a straightforward
way.   For a loop with variable cross-section the energy equation (eq.~[\ref{eq:balance}]) takes the form
\be
 \frac{\kappa_{0}}{A(s)} \frac{d}{dz}(A(s)T^{5/2}\frac{dT}{dz})\,+
\,E_{h}\,-\,P_{0}^2\chi_{0}T^{-(2+\gamma)}\,=\,0,
\label{eq:energy-diameter}
\ee
where $A(s)$ represents the variable cross-section as a function of distance along the axis
of the elementary strand.  To make this equation amenable to analytical treatment  Following Bray
et al.\ (1991, p.\ 294) I assume a power-law relation between cross-section and temperature,
\be
A(s)\, =\, A_0(T(s)/T_{max})^{\delta},
\label{eq:cross-section}
\ee
where $A(0)$ is the cross-section at the apex of the strand, and $\delta$ is a parameter.
Note that for a strand that expands with height $\delta$ is positive, and that constant cross-section
corresponds to $\delta$ = 0.

Using the same technique as in \S\,~\ref{sec:formalism} to reduce the energy equation to its
dimensionless form I find the remarkable result that the energy equation in terms of the variable $\eta$
(eq.~[\ref{eq:eta-de}]), with parameters $\nu$ and $\mu$ remains unchanged.  The expression for 
$\lambda$ in terms of $\nu$ and $\mu$, equation~(\ref{eq:lambda1}), as well as the boundary conditions,
remain unchanged as well, while the definition of the parameters $\nu$ and $\mu$ becomes
\begin{eqnarray}
\nu\,=\, \frac{\alpha+\delta}{7/2+\delta},\\
\mu\,=\, \frac{\delta-(2+\gamma)}{\delta+7/2}.
\end{eqnarray}
The main difference with previous results is in the definition of the dimensionless temperature,
\be
t\, =\, \eta^{\frac{2}{7+2\delta}},
\ee
which results in only a small correction in the expression for the temperature profile of the strand:
\be
T(s)\, =\, T_{max}[\beta_n^{-1}(s/L;\lambda+1,1/2)]^{\frac{1}{(1+2\delta/7)(2+\gamma+\alpha)}}.
\label{eq:t-delta}
\ee
The other differences are in the expressions for $\xi$, and $\epsilon$.  After some straightforward
algebra, the explicit expressions for the scaling laws of tapered flux tubes become
\begin{eqnarray}
E_{top}\, =\, \frac{P_0^2\chi_0(7/2+\alpha)} {T_{max}^{2+\gamma}(3/2-\gamma+2\delta)}, \\
P_{0}L\, =\, T_{max}^{\frac{11+2\gamma}{4}}\, [\frac{\kappa_{0}}{\chi_{0}}]^{1/2}\, 
\frac{(3-2\gamma+2\delta)^{1/2}}{4+2\gamma+2\alpha}\, B(\lambda+1,1/2).
\label{eq:tapered-scaling}
\end{eqnarray}
Without solving for the temperature profile Bray et al.\ (1991, p.\ 297) have derived the same
results for the scaling laws of tapered loops.

\begin{figure*}[!t]
\epsscale{2.20}
\plottwo{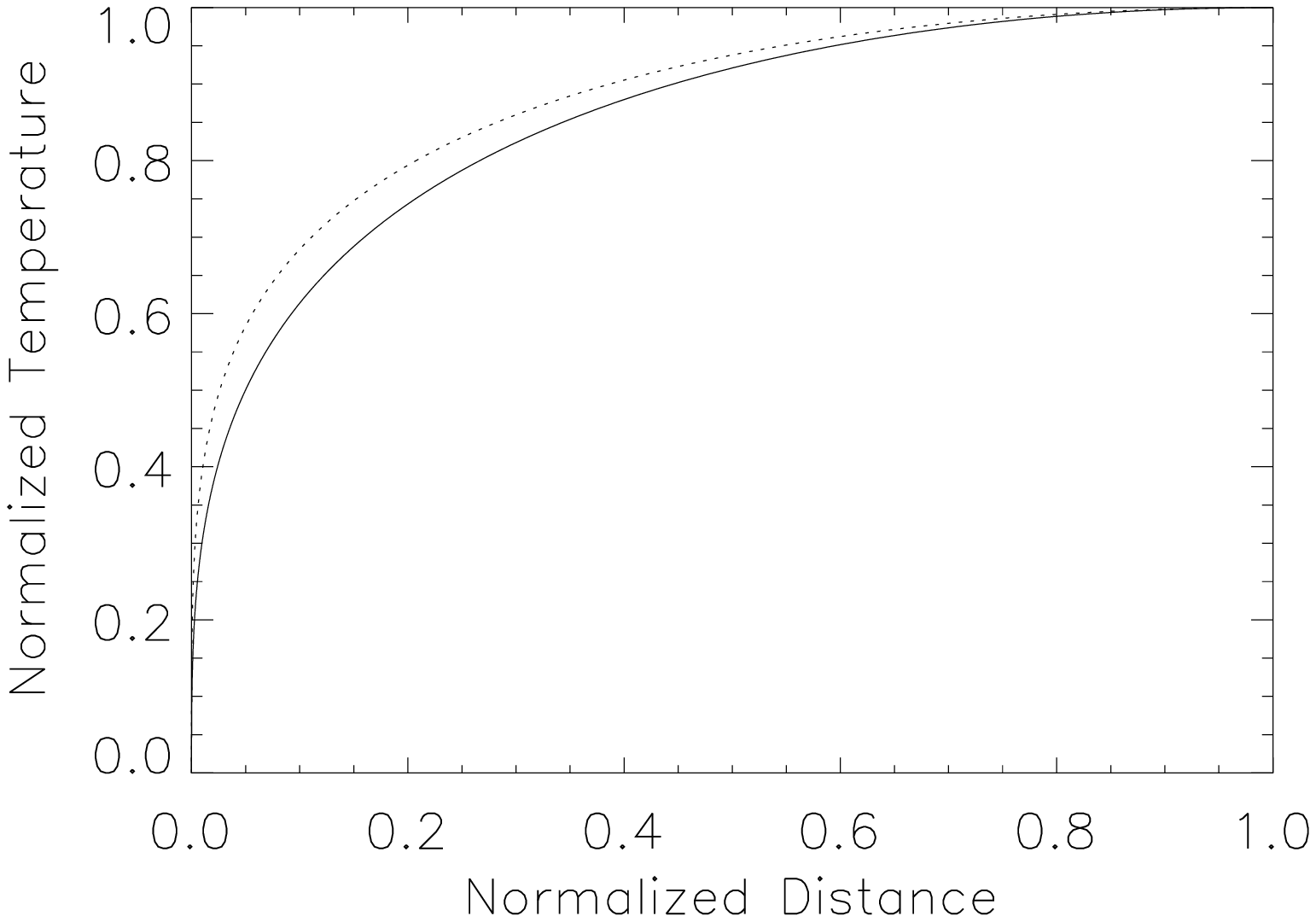}{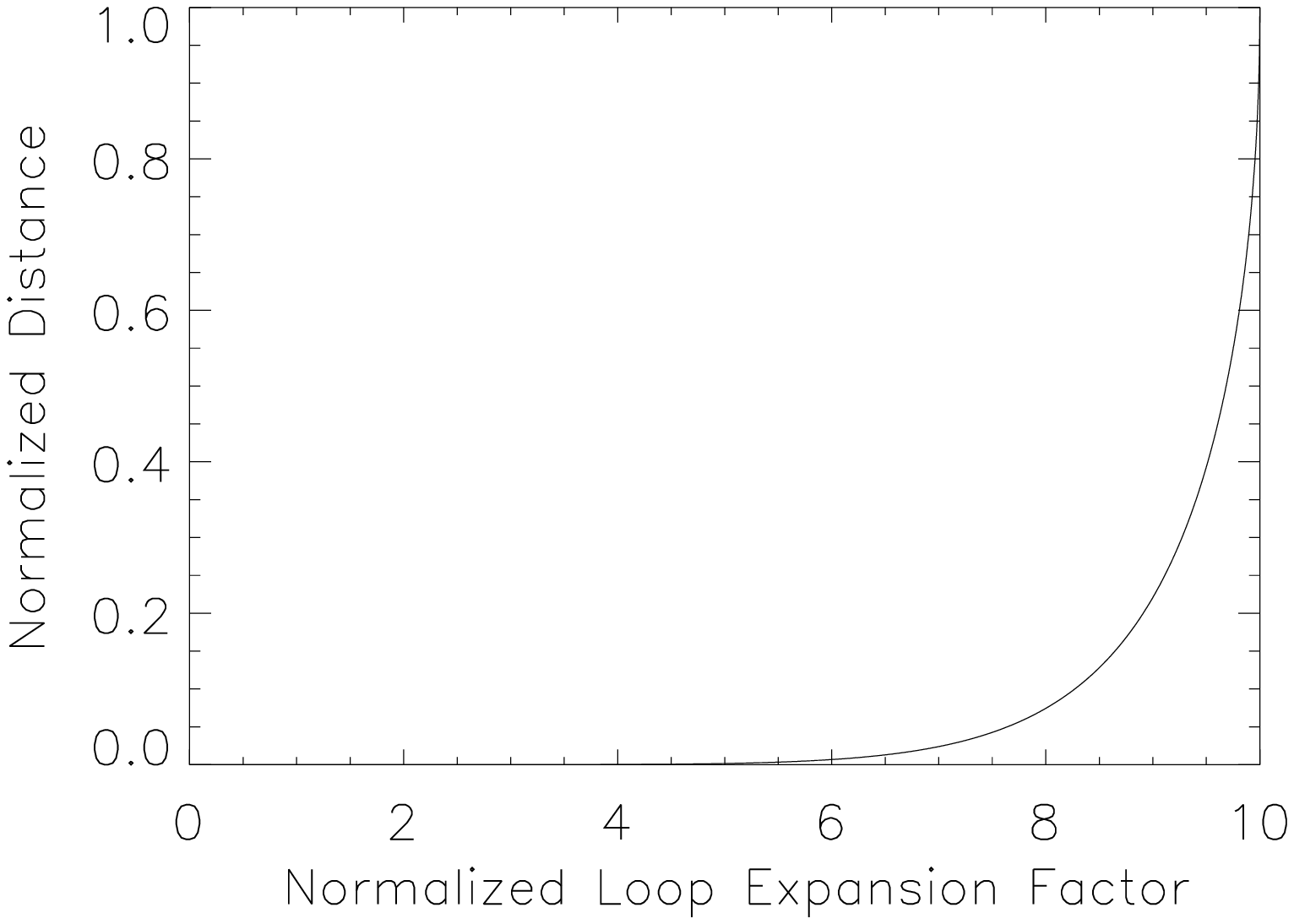}
\caption{Left:  Dimensionless temperature solutions for an Ohmically heated
($E_h\, \thicksim\,T^{-3/2}$) elementary strand with constant cross-section
($\delta$=0, thin line), and an increase in strand radius proportional to $\sqrt{T}$
($\delta$=1, dashed line).
Right:  Dimensionless loop radius as a function of distance along the loop axis for
the expanding loop.
\label{fig:expansion}}
\end{figure*}

The difference of the expressions above with the scaling laws for constant cross-section amounts
to a correction factor 1/(1+2$\delta$/7) in the power-law relation between $t$ and $\eta$
for the temperature profile (eq.~[\ref{eq:t-delta}]), and a minor correction in the second
scaling law.  However, in the first scaling law the correction can be significant.  A value
of $\delta$~=~1, which as will be shown below is not extravagant, leads to an over-pressure
of a factor $\sqrt{3} \thicksim$~1.7, which is of the same order as the over-pressure that can be
generated by footpoint heating in comparison with uniform heating.   This complicates matters, but
I will show below that potentially the integrated brightness of ``moss" can be used to determine the
amount of loop tapering independently.

Figure~\ref{fig:expansion} illustrates the effect of flux tube expansion on the solutions, for the case of
current heating with Spitzer resistivity, discussed above ($\alpha$=-3/2).  The left panel shows the
temperature profile for both constant cross-section ($\delta$=0) and a loop expanding in radius as the
square root of temperature ($\delta$=1, dashed curve), and the right panel shows the loop radius as a function
of distance along the axis from bottom to top.  The assumed temperature dependence of the expansion rate 
produces roughly a factor ten ratio in loop radius between the apex and the chromospheric footpoint of the 
strand, in  agreement with the expansion factors that follow from field extrapolations (see DeForest, 2007).

These results show that here is only a relatively small difference in the temperature profile and pressure
between  strands with uniform and expanding cross-sections.  Therefore the ratio in loop integrated 
emission measure between
a section of the loop near the top and one near the footpoints constitutes a good diagnostic for the 
expansion factor, since the emission scales linearly with the cross-sectional area.  At the same time, for
well resolved strands the emission per pixel will barely be influenced by the expansion factor.

Note that strand expansion may have a significant influence on the emission from moss, footpoint
EUV emission from hot loops (Peres, Reale, \& Golub 1994; Schrijver et al.\ 1999;
Berger et al.\ 1999; Fletcher \& DePontieu 1999).
It was pointed out by Martens et al.\ (2000) that the moss emission can be used as a diagnostic for
the strand footpoint constriction.
\nocite{schrijver-etal99, berger-etal99, fletcher-depontieu99, martens-kankelborg-berger00,
peres-reale-golub94}
In the example of Figure~\ref{fig:expansion} a hot strand with a maximum temperature of 3.5 MK, will
have its cross-section area constrained by a factor 0.29 at 1 MK, the temperature that dominates the
emission in the {\it TRACE} 171 \AA\ channel, which shows the moss very well
(e.g., see Fig.~\ref{fig:euv-sxr}).   
Since according to Figure~\ref{fig:expansion} there is little change in the temperature gradient at
that point, the emission measure of the moss is also reduced by a factor of approximately 0.3 when
compared to a loop of constant cross-section.

Fletcher \& DePontieu (1999) and Martens et al.\ (2000), both using {\it TRACE} data, as well as recently
Warren et al.\ (2007, private communication) and Winebarger, Warren, \& Falconer (2008), using
the  {\it Hinode} EUV Imaging Spectrometer (EIS) find a filling factor for
\nocite{winebarger-warren-falconer08}
the moss of the order of 10\%, less than the result given above.  As discussed in Martens et al.\  the 
filling factor of the moss is the product of the filling factor in the coronal portion of the loop times
the magnetic constriction near the loop footpoints.   The calculation above only considers the
magnetic constriction factor, and combined with a coronal filling factor also of the order of 0.3 would
lead to the observed result of about 10\%.  Independent observations of the filling factor in the
coronal portion of the loops are needed to separate the two components of the filling factor.
For the purpose of estimating the constriction factor $\delta$, as discussed above, only the ratio
between the loop apex emission measure and that of the moss needs to be known.

\subsection{Solutions with Temperature Reversals:  Physical Interpretation \label{sec:multiple-t}}

In \S\,~\ref{sec:nonuniform} the existence of solutions with temperature reversals has been
demonstrated.  The question arises whether such solutions can occur in physical reality, and I
will demonstrate here that in certain circumstances the answer may be positive.  

In the simplest solution with a temperature reversal ($n$=2) one would find a condensation with
a chromospheric temperature at the top of the loop.  Such solutions have been investigated
numerically by Karpen et al.\  (2001, 2003, 2005, 2006), 
\nocite{karpen-etal01, karpen-etal03, karpen-etal05, karpen-antiochos-klimchuk06}
and these authors found that they are
unstable in the usual semi-circular loops because of gravity:   the cool ``blob" will be pulled down,
in a similar fashion as in the Rayleigh-Taylor instability.   It is reasonable to expect the same
effect for higher order solutions with multiple cool blobs.

However, in order to investigate plausible geometries for solar prominences, Karpen et al.\  further
simulate strands that are stretched out in the horizontal direction (i.e., loops that are mostly
horizontal, except near both footpoints),  and also loops that  have a dip (a U-shaped section)
in the middle, a widely accepted physical model for sustaining prominence material in the
corona far above its scale-height.  They find that such situations can be thermally stable in some
cases, or are maintained for relatively long times before dissipating in others.  I suggest that these 
numerical solutions are a manifestation of the temperature reversal solutions derived in this paper.

The following thought experiment further verifies the physical viability of the temperature
reversal solutions.  The solution with a uniformly increasing temperature stretches from one
footpoint to the top of the loop.  In the other half of the loop the solution must be the same,
i.e., same $T_{max}$, loop half-length $L$, and pressure $P_0$, but mirrored in the space
coordinate of course.  That situation is physically identical to the $n$=2 solution for a loop
with the same maximum temperature but twice the loop length.  In particular the loop
pressure should be the same.

By writing out the eigenvalue solution for $\epsilon_n$ (eq.~[\ref{eq:epsilon-n-scaling}]) in
terms of the original physical parameters one finds the scaling law(s) for loops with
temperature reversals,
\be
P_{0}L/n\, =\, T_{max}^{\frac{11+2\gamma}{4}}\, [\frac{\kappa_{0}}{\chi_{0}}]^{1/2}\, 
\frac{(3-2\gamma)^{1/2}}{4+2\gamma+2\alpha}\, B(\lambda+1,1/2).
\label{eq:multiple-t-scaling}
\ee
This shows that the $n$=2 solution for the loop with twice the loop-length as the original one,
and with the same temperature and heating profile, indeed produces the same loop pressure,
as the thought experiment requires.  This scaling law also demonstrates that for the same
loop with the same heating profile and maximum temperature, the loop pressure in the
case of temperature reversal profiles is an $n$-multiple of the pressure of that in the loop
without condensations.

I conclude from this section that it is plausible that the formalism of this paper captures not
only the temperature profile of ordinary hot coronal loops, but also, in a crude sense, that of
solar prominences carrying condensations.

\section{Discussion \label{sec:discussion}}

Based upon the observations that emission from the solar corona is dominated by
Active Region soft X-ray loops, and that these loops are mostly stable for times
longer than their thermal timescales (e.g., Antiochos et al.\ 2003), and coupled with
the likelihood that observed coronal loops consist of a multitude of thermally
isolated strands, I have set out to derive analytical models for the thermal structure
of quasi-static and isobaric coronal strands.
I have derived analytical solutions for the temperature structure and scaling laws
in coronal strands in a formalism for the coronal heating profile that captures loop
footpoint heating, uniform heating, as well as loop apex heating.  Solutions with both
uniformly increasing temperature and with temperature reversals have been derived,
and it is suggested, from a comparison with numerically derived results in Karpen et al.\
(2006, and references therein) that the latter solutions apply to the temperature profile
in prominences.

The analytical solutions produce strand temperature profiles that vary very little for
enormous differences in the distribution of the heating over the loop, thus making the
temperature profile a rather poor diagnostic for the heating functions.  On the other
hand, I demonstrated that the constant of proportionality in the RTV scaling
law ($P_0L\,\thicksim\,T_{max}^3$) depends on the distribution of the heating over the
strand, with footpoint heating producing over-pressure in comparison to uniform heating. 
Hence, measurements of pressure or density, for instance from density sensitive line ratios,
in combination with temperature measurements may be very useful in teasing out the
distribution of the heating over observed loops.

The analytical solutions have been derived for strands with uniform cross-section, but
a generalization to a family of solutions for non-uniform cross-sections has been
demonstrated as well, and it turns out that these solutions only differ only by a small correction
factor from the solutions for constant strand cross-section.  Through a simple example
it is demonstrated that variation in strand-cross section can account for the deficit in
observed ``moss" emission in comparison with the models.  As has already been pointed
out by DeForest (2007) the same effect can account for the observed difference between
scale-height temperature and observed temperature in EUV loops.

I have further shown that the expression used for the coronal loop heating function is quite
versatile, by taking a quintet of well known leading candidates for the coronal heating
mechanism and explicitly expressing them in terms of the heating formalism.  However,
I should point out here that certainly not all proposed coronal heating mechanisms can
be expressed in the form $E_{h}\,\thicksim\, P_{0}^{\beta}T^{\alpha}$.  For example, a
wave flux entering a strand from the chromosphere, and dissipating on a length-scale
$\ell$ that is short compared to the loop-length, will lead to a heating profile of a
form that is explicitly dependent on the distance from the footpoint $s$, e.g., $E_h\,
\thicksim\, \exp{(-s/\ell)}$, and that therefore cannot be expressed in terms of
local thermodynamic variables alone.  Such an expression has been used in numerical
simulations by a number of authors, for instance Serio et al.\ (1981), Aschwanden et
al.\ (2001), and M\"{u}ller et al. (2005).  Their numerical solutions exhibit a variety of
features that are very similar to those of the analytical solutions derived here:  there is
over-pressure in loops with footpoint heating, and when the heating is concentrated
too strongly near the footpoints the quasi-static solution ceases to exist (M\"{u}ller et
al.\  2005), just as I find for $\alpha\,<\,-2.5$.

The temperature profiles presented here are expressed in terms of the inverse of the
incomplete regularized $\beta$-function.  This type of solution has first been derived
by Martens (1981) for application in stellar coronae, and has been applied to solar
coronal loops by Kuin \& Martens (1982) for the case of uniform heating. 
\nocite{martens81, kuin-martens82} 
The solution has been used extensively by  for the purpose of solar and stellar loop
modeling by Fisher \& Hawley (1990) and Hawley \& Fisher (1992, 1994).
\nocite{fisher-hawley90, hawley-fisher92, hawley-fisher94}
To my knowledge the analytical solutions for the temperature profile expressed as
incomplete $\beta$-functions in the current paper, are the first for non-uniform
heating profiles.
As pointed out above, Craig et al.\ (1978) were the first to show that the form of the
first scaling law is preserved for non-uniform heating expressed in terms of pressure
and density to a given power (eq.~[\ref{eq:heatingf}]).  Bray et al.\ (1991) first
calculated the heating dependent constant of proportionality.
Kano \& Tsuneta (1995, 1996) added a non-zero incoming thermal conductive flux
at the loop apex to model the apparent loop top heating in flaring loops and derived
expressions for the scaling laws in terms of inequalities as well as a scaling law
for loops heated completely at the apex.

As mentioned in the introduction, having analytical solutions for the temperature
profile and pressure in loop strands for a number of leading candidate coronal heating
mechanisms makes it a lot simpler to calculate DEMs 
per pixel in forward modeling of coronal loops consisting of a multitude of unresolved
strands.  This kind of capability will be most useful for analyzing coronal loops with the
AIA instrument onboard SDO that will observe coronal loops through six narrowband
EUV bandpasses, in combination with XRT on {\it Hinode} that has nine broadband
filters for observing soft X-ray emission.
However, before this capability can be used one has to investigate the {\it existence,
stability,} and {\it accuracy} of the analytical solutions derived here.  In particular
one has to analyze the effects of the assumptions of constant pressure, zero
temperature and flux at the lower boundary, and the approximation of the radiative
loss function by a single power-law.  For the case
of uniform heating ($\alpha$=$\beta$=0 in the heating formalism of this paper),
the existence and stability of the solutions has been confirmed many times over
by numerical simulations.  Martens et al.\ (2000) also considered the accuracy of
this solution by comparing the numerical and analytical solution, and they found a
temperature difference of a few percent at most at every location in the loop
for loop heights smaller than the pressure scale height for the apex temperature. 

The numerical analysis of the solutions for non-uniform heating is in progress
(Martens, Winter, \& Munetsi-Mugomba 2007) and has yielded so far that for
$\alpha\,>\,-2$  \nocite{martens-winter-munetsi-mugomba07}
stable static solutions exist, and that are very accurate, as long as the loop height
is less than the apex temperature pressure scale height.  For heating that is strongly
concentrated near the footpoints we found cyclic solutions very similar to those
obtained by M\"{u}ller et al.\ (2005) (and anticipated by Kuin \& Martens, 1982) for
very small values of the dissipation length
$\ell$ (see above) compared to the loop length.  These results will be
demonstrated and analyzed in much more detail in an upcoming paper
(Winter, Martens, \& Munetsi-Mugomba 2008, in preparation).
I note that I have found the study of stability through an analytical linear stability
analysis too cumbersome, and inevitably leading to a numerical analysis anyway,
which is why we attacked the problem with a full time-dependendent numerical
simulation from the onset.

In conclusion then, I have generalized the analytical solutions for the scaling
laws and temperature profile first published by Rosner et al.\ (1978b) from the
case of uniform heating to a family of non-uniform heating functions, including
footpoint and loop apex heating.   These results lead to increased insight in the
relations between the thermodynamic variables and heating parameters in loop
strands, and will substantially reduce the effort required for forward modeling
of pixel Differential Emission Measures of multi-stranded loops to be compared
with combined data from XRT and AIA.

\acknowledgments

This research has made use of NASA's Astrophysics Data System and the
Virtual Solar Observatory.  It has
been supported with {\it TRACE}, {\it Hinode}-XRT, and SDO/AIA funds.
{\it Hinode} is a Japanese mission developed and launched by ISAS/JAXA, with
NAOJ as domestic partner and NASA and STFC (UK) as international partners.
It is operated by these agencies in co-operation with ESA and the NSC (Norway).
{\it TRACE} mission operations and data analysis are supported by NASA Grant
NAS5-38099 to Lockheed-Martin Solar and Astrophysics Laboratory, with
subcontracts to Montana State University and the Harvard-Smithsonian
Center for Astrophysics.
AIA mission preparation at the Harvard-Smithsonian Center for Astrophysics and
Montana State University is supported by NASA Grant NNG04EA00C to
Lockheed-Martin Solar and Astrophysics Laboratory with  subcontracts to the
Smithsonian Astrophysical Observatory and Montana State University.
I am grateful to Dr.\ Ryouhei Kano (NAOJ) for constructive comments on a draft
version of this paper.




\appendix

\section{Integration for Non-zero Chromospheric Temperature}

The dimensionless integral expression for the temperature profile in
equation~(\ref{eq:integrand}) in \S~\ref{sec:t-solution} is
\be
x\, =\, \frac{[\epsilon(\mu+1)/2]^{1/2}}{\nu-\mu} \int_{0}^{u}\frac{z^{\lambda}dz}{(1-z)^{1/2}},
\label{eq:A1}
\ee
with the parameters defined in that same section.  When the temperature at the
chromospheric boundary at $x$~=~0 is not ignored, we a have lower boundary
condition $u_0$ defined by $u_0\,=\,
 [\frac{T_{chrom}}{T_{max}}]^{2+\alpha+\gamma}$  which is a very small number
except for $\alpha\, \downarrow\, -2.5$.

Equation~(\ref{eq:A1}) becomes
\be
x\, =\, \frac{[\epsilon(\mu+1)/2]^{1/2}}{\nu-\mu}
[ \int_{0}^{u}\frac{z^{\lambda}dz}{(1-z)^{1/2}}\, -\, 
\int_{0}^{u_0}\frac{z^{\lambda}dz}{(1-z)^{1/2}}],
\label{eq:A2}
\ee
or, after integration
\be
x\, =\, [\frac{\epsilon(1-2(2+\gamma)/7)}{2}]^{1/2}\, \frac{7
[\beta(u;\lambda+1,1/2)-\beta(u_0;\lambda+1,1/2)]}{2(2+\gamma+\alpha)}.
\label{eq:A3}
\ee
For the same value of $\epsilon$ the extra term would merely amount to a
small shift to the left along the x-axis for the solution.  However, because the
boundary condition $u\, =\, 1$ at $x\, =\, 1$ also has to be met, the value
of $\epsilon$ -- and thereby the constant in the first scaling law -- will
change.

Applying the boundary conditions at the top of the loop yields
\be
1\, = \, [\frac{\epsilon(1-2(2+\gamma)/7)}{2}]^{1/2}\, \frac{7[B(\lambda+1,1/2)
-\beta(u_0;\lambda+1,1/2)]}{2(2+\gamma+\alpha)},
\label{eq:A4}
\ee
where again B is the complete $\beta$ function.  The value of the term
expressed by the incomplete $\beta$-function is small compared to that
of the complete $\beta$-function, B($\lambda$+1,1/2) for $\lambda\,>$~-1,
i.e., $\alpha$~$>$~-2.5, as will be shown presently.

For small $u_0$ the denominator in the integrand term of the incomplete
$\beta$-functioncan be set to unity, i.e.,
\be
\beta(u_0;\lambda+1,1/2)\,=\,\int_{0}^{u_0}\frac{z^{\lambda}dz}{(1-z)^{1/2}}\,
\approx\,\int_{0}^{u_0}z^{\lambda}dz\,=\,\frac{{u_0}^{\lambda+1}}{\lambda+1}.
\label{eq:A5}
\ee
Restoring the original parameters $t_0$, $\alpha$, and $\gamma$ (see
\S\,~\ref{sec:t-solution}) yields
\be
\frac{{u_0}^{\lambda+1}}{\lambda+1}\,=\,\frac{(2+\alpha+\gamma)
t_0^{11/4+\gamma/2}}{11/4+\gamma/2},
\label{eq:A6}
\ee
which is of the order of 10$^{-6}$ for $t_0\,\thicksim\,10^{-2}$, while the
value of the complete $\beta$-function is of the order 0.1 for the allowed
positive values of $\lambda$.  Note that
for $\alpha\,\downarrow\,-2.5$, $u_0$ is no longer very small, even while
$t_0$ is, but despite of that the value of the incomplete $\beta$-function
remains small, because the value of $\lambda$ in the integral expression
of equation~(\ref{eq:A5}) becomes very large.  This has been verified
numerically.

Hence equation~(\ref{eq:A4}) produces a correction of  the order of 0.3\%
for the value of $\epsilon$, and therefrom a correction of a fraction of again
10$^{-5}$ in the loop pressure according to equation~(\ref{eq:first-scaling}),
a correction that I have ignored in the body of the paper.


\bibliographystyle{apj}
\bibliography{myrefs}

\end{document}